\documentstyle[11pt]{article}
\begin{document}

\setcounter{page}{0}

\begin{flushright}
CERN-TH. 7440/94
\end{flushright}
\vspace{1.0cm}

\begin{center}
{\LARGE
{\bf EFFECTIVE FIELD THEORY APPROACH}

{\bf TO PARTON-HADRON CONVERSION}
\vspace{0.5cm}

{\bf IN HIGH ENERGY QCD PROCESSES}
}
\end{center}
\bigskip

\begin{center}

{\Large
{\bf K. Geiger}
}

{\it CERN TH-Division, CH-1211 Geneva 23, Switzerland}
\end{center}
\vspace{2.0cm}

\begin{center}
{\large {\bf Abstract}}
\end{center}
\medskip

A QCD based  effective action is constructed to describe the dynamics
of confinement and symmetry breaking in the process
of parton-hadron conversion.  The deconfined quark and gluon degrees of freedom
of the perturbative QCD vacuum are coupled to color singlet collective fields
representing
the non-perturbative vacuum with broken scale and chiral symmetry.
The effective action recovers QCD with its scale
and chiral symmetry properties at short space-time distances, but yields
at large distances ($r\,\lower3pt\hbox{$\buildrel > \over\sim$}\,1$ fm)
to the formation of symmetry breaking gluon and quark condensates.
The approach is applied to the evolution of a
fragmenting $q\bar q$ pair with its generated gluon distribution, starting from
a large hard scale $Q^2$.
The modification of the gluon distribution arising from the coupling to the
non-perturbative
collective field results eventually in a complete condensation of gluons.
Color flux tube configurations of the gluons in between the $q\bar q$ pair are
obtained
as solutions of the equations of motion. With reasonable parameter choice, the
associated energy per unit length (string tension) comes out $\simeq 1$ GeV/fm,
consistent with  common estimates.
\noindent

\vspace{0.5cm}

\rightline{e-mail: klaus@surya11.cern.ch}
\leftline{CERN-TH. 7440/94,  September 1994}

\newpage

\noindent {\bf 1. INTRODUCTION}
\bigskip

The physics of QCD exhibits different relevant excitations at
different length scales. At space-time short distances (below 1 fm)
the relevant degrees of freedom are quarks and gluons whose
interactions are well described by perturbative QCD \cite{pQCD1,pQCD2}.
The long distance physics on the other hand is governed by the
hadronic degrees of freedom, and
the particles which are observed at large scales are hadrons
whose interactions are well described by chiral models \cite{chiral}.
The change of resolution of our microscope with which we probe
the physics of QCD is formally described by a renormalization group
equation, or evolution equation, that determines the scale dependence
of the theory \cite{polchinski84,reuter93,bonini93}.

The transition from short distance (high momentum transfer) regime
to the long distance (low energy) domain can be cast in terms of an
evolution equation for an {\it effective QCD action} that embodies {\it both},
fundamental
partonic degrees and hadronic degrees of freedom. By increasing
the distance scale (decreasing the momentum scale), the evolution
\cite{ellwanger94} of the effective field theory must lead from one set to the
other set
of degrees of freedom.

Experiments on high energy QCD processes, such as
$e^+e^-$ annihilation, deep inelastic $ep$ scattering, Drell Yan, etc.,
strongly support the conception that the observed parton fragmentation into
hadrons is a universal mechanism. Moreover,
the dynamical transformation of color charged quarks and gluons in high energy
QCD processes into colorless hadrons is commonly believed to
be a local phenomenon \cite{lphd}. Thus, a consistent description of the
local hadronization mechanism must be independent of the details
of the partons prehistory and should in principle apply also to
hadron-hadron, hadron-nucleus, or nucleus-nucleus collisions.

To date most of the theoretical tools to study properties of QCD are
inadequate to describe the dynamics of the transformation from
partonic to hadronic degrees of freedom:
Perturbative techniques are limited to the deconfined, short
distance regime of high energy partons \cite{dok80}, QCD sumrules
\cite{shifman79} and effective low energy models \cite{witten83} are
restricted to the long distance domain of hadrons, and
lattice QCD \cite{karsch89} lacks the capability of dynamical
calculations concerning the quark-gluon to hadron conversion.
On the other hand, phenomenological approaches to parton
fragmentation \cite{webber86}, are mostly based on
hadronization models with adhoc prescriptions to simulate
hadron formation from parton decays.

In this paper I follow a rather different, universal approach
to the dynamic transition {\it between} partons and hadrons based on an
effective
QCD field theory description, as recently proposed in Ref. \cite{ms33}.
In the spirit of the aforementioned evolution of effective field theory
from high energy to low energy scales \cite{reuter93,ellwanger94},
the key element is to project out
the {\it relevant degrees of freedom} for each kinematic regime and to embody
them in an effective QCD Lagrangian which recovers QCD with its scale
and chiral symmetry properties at high momentum transfer, but yields
at low energies the formation of symmetry breaking gluon and
quark condensates including excitations that represent
the physical hadrons.
In Sec. 2, I will first formulate the general field theoretical
framework. On the basis of  the dual vacuum picture of coexisting
perturbative and non-perturbative domains an effective action is constructed
that embodies the correct scale and symmetry properties of QCD.
The concept is here more phenomenologically motivated than the related
formal approach of Ref. \cite{ellwanger94}. However, there appears to  be a
clear
correspondance bertween those two descriptions.
In Sec. 3, I shall demonstrate the applicability of this effective QCD field
theory
to the dynamics of parton-hadron conversion by exemplarily
considering the evolution of gluons produced by a fragmenting quark-antiquark
pair.
The change of the gluon distribution in the presence of
a confining composite field is studied and flux tube solutions of the gluon
field
resulting from the equations of motion are analyzed in terms
of the string tension, that characterizes the effective confinement potential.
Various perspectives of the approach are discussed in Sec. 4, in particular the
applicability to
the QCD phase transition, and high density QCD.
\bigskip

\noindent {\bf 2. EFFECTIVE QCD FIELD THEORY WITH SPONTANOUS SYMMETRY BREAKING}
\bigskip

The goal is to construct an effective field theory that describes the dynamics
of {\it both} partonic and hadronic degrees of freedom and their interplay.
The approach is based on the concept of the effective
action \cite{polchinski84,ellwanger94}, which will be represented here as
($r\equiv r^\mu$ denotes the space-time 4-vector)
\begin{equation}
S_{eff}\;=\; \int d^4 r \,\left[
\frac{}{}
\cal{L}[\psi,A] \;+\; \cal{L}[\chi,U]  \;+\; \cal{L}[\psi,A,\chi]
\right]
\;\;.
\label{S}
\end{equation}
The three contributions to the action,
which will be discussed below, correspond to the QCD Lagrangian
with the quark ($\psi_i$) and gluon fields ($A_a$),  an effective
low energy Lagrangian introducing  composite fields $\chi$ and $U$, and
a term that couples the "microscopic" fundamental quark and gluon degrees of
freedom to the "macroscopic" fields $\chi$ and $U$ which represent the
hadronic degrees of freedom.
\bigskip

\noindent {\bf 2.1 {\boldmath $\cal{L}[\psi,A]$}}
\medskip

The QCD Lagrangian in (\ref{S}) contains the gluon fields $A_a^\mu$
coupled to massless quark fields $\psi_i$ ($i=1,\ldots , N_f$),
\begin{equation}
\cal{L}[\psi,A] \;=\; -\frac{1}{4}\,\,F_{\mu\nu, a} F^{\mu\nu}_a
\;+\;  \overline{\psi}_i \left[\frac{}{}\,i \gamma_\mu \partial ^\mu \,
- g_s  \gamma_\mu A^\mu_a T_a  \right]\, \psi_i
\;+\;
\cal{L}_{gauge} \;+\; \cal{L}_{ghost}
\label{LQCD}
\;.
\end{equation}
Here
$F_a^{\mu\nu}= \partial^\mu A_a^\nu -\partial^\nu A_a^\mu + g_s f_{abc} A^\mu_b
A^\nu_c$
is the gluon field strength tensor. The subscripts $a, b, c$ label the
color components and $g_s$ denotes the color charge related to  $\alpha_s
=g_s^2/(4\pi)$.
The $T_a$ are the generators of the $SU(3)$ color group, satisfying
$[T_a,T_b] = i f_{abc} T_c$ with the structure constants $f_{abc}$.
The gauge fixing term
$\cal{L}_{gauge}=1/(2 a) (\eta_\mu A_a^\mu)^2$
with gauge parameters $a$ and $\eta_\mu$, and
the contribution of Fadeev-Popov ghost fields $\zeta$,
$\cal{L}_{ghost} = (\partial_\mu \zeta _a^\ast )(\delta_{ab} \partial^\mu - g_s
f_{abc} A_c^\mu) \zeta_b$,
will be irrelevant later on, because a physical
gauge $\eta \cdot A =0$ can be fixed, which eliminates the presence of ghosts.

The Lagrangian (\ref{LQCD}) is well known to be invariant
under chiral transformations \cite{chiral}.
At the tree level it is also invariant under scale
transformations $r_\mu \rightarrow r_\mu' = e^\Delta r_\mu$ \cite{ellis70},
generated by a so-called dilaton charge $D(t)=\int d^3r J^S_0(r)$, where
$J_\mu^{S}$ is
the scale current and
$\left[\frac{}{}D , \varphi(r)\right]= i \left( r_\mu \partial ^\mu
+d_\varphi\right) \varphi (r)$
for a generic quantum field $\varphi$ with scale dimension $d_\varphi$.
The convention is $d_A =1$ for gauge boson fields and $d_\psi = 3/2$ for
fermion fields.
It follows that
$\cal{L}(\psi,A)$
has scale dimension 4 so that
$\left[\frac{}{}D , \cal{L}(\psi,A)\right]=0$
and therefore massless QCD proves to be scale invariant at tree level.

At high energies and short space-time distances, asymptotic freedom
leads to unconfined gluon and quark fields in (\ref{LQCD}).
However, in the physical world
these color degrees of freedom are confined, and both chiral and
scale symmetry are explicitely broken.
To describe  the dynamics of the symmetry breakdown
of the transition between the perturbative, scale and chiral invariant,
regime and the non-perturbative world with broken symmetries,
one needs to supplement  (\ref{LQCD}) (by adding
$\Delta \cal{L}=\cal{L}[\chi,U]+\cal{L}[\psi,A,\chi]$)
to construct an effective description
such that at high energies the fully symmetric QCD phase is recovered,
but at low energies massive hadrons emerge.
\newpage

\noindent {\bf 2.2 {\boldmath  $\cal{L}[\chi,U]$}}
\medskip

The specific form of $\cal{L}[\chi,U]$ in (\ref{S})
is adopted from Refs. \cite{janik84,ellis90}, where
an effective low energy Lagrangian was constructed, guided by
the scale and chiral symmetry properties of the
QCD Lagrangian.
The construction
is based on the observation that even massless QCD is no longer
scale invariant when going beyond tree level, because of the
scale anomaly \cite{ellis70}
$\partial ^\mu J_\mu^S= \beta(g_s)/(2 g_s)  F_{\mu\nu} F^{\mu\nu}$,
where $J_\mu^S$ is the scale current as before.
In addition chiral invariance breaks down when finite quark masses are
taken into account.
As a consequence, the QCD energy momentum tensor $\theta_{\mu\nu}$
exhibits the well known {\it trace anomaly} \cite{shifman91}:
\begin{equation}
\theta_\mu^\mu\;=\;\frac{\beta(\alpha_s)}{4 \alpha_s} \,F_{\mu\nu}F^{\mu\nu}
\;+\;(1 + \gamma_m)\;\sum_q \,m_q \bar q q
\label{theta}
\;,
\end{equation}
where $\beta$ is the Callan-Symanzik function and $\gamma_m$ is the
anomalous mass dimension.
This anomaly constrains the form of the effective low energy Lagrangian,
because
without it Poincar\'e invariance would be broken, and the mass of
the proton would come out wrong, since $2 m_p^2 = \langle
p|\theta_\mu^\mu|p\rangle$.

The extension of these symmetry properties of the QCD Lagrangian to the
low energy domain was modeled by Campbell, Ellis and Olive \cite{ellis90}
as\footnote{In Ref. \cite{ellis90} the second term in eq. (\ref{Lchi}) was
scaled by a factor
$(\chi/\chi_0)^2$, which however is not necessary \cite{janik84}, since just
this term
gives a chiral symmetric contribution in agreement with the QCD anomaly
(\ref{theta}).}
\begin{eqnarray}
\cal{L}[\chi,U] &=&
\frac{1}{2}\,(\partial_\mu \chi) ( \partial^\mu \chi )
\;+\; \frac{1}{4}\,Tr\left[\frac{}{}(\partial_\mu U)
( \partial^\mu U^\dagger ) \right]
\;-\;
b\, \left[\frac{1}{4} \,\chi_0^4 \,+\,\chi^4\,
\ln\left(\frac{\chi}{e^{1/4} \chi_0}\right)\right]
\nonumber \\
& &
-\;c\; Tr\left[\frac{}{}  m_q (U + U^\dagger) \right]
\;\left(\frac{\chi}{\chi_0}\right)^3 \;
\;-\;\frac{1}{2} \,m_0^2\,\phi_0^2 \; \left(\frac{\chi}{\chi_0}\right)^4
\label{Lchi}
\;.
\end{eqnarray}
This form
introduces a {\it scalar gluon condensate field} $\chi$ and
a {\it pseudoscalar quark condensate field}
$U = f_\pi \exp\left(i \sum_{j=0}^8 \lambda_j\phi_j/f_\pi\right)$
for the nonet of the meson fields $\phi_j$
($f_\pi=93$ MeV, $Tr[\lambda_i\lambda_j]=2 \delta_{ij}$, $U
U^\dagger=f_\pi^2$), with
non-vanishing vacuum expectation values
\begin{equation}
\chi_0 \;=\;\langle\, 0\,|\;\chi\; |\,0\,\rangle\; \ne \;0
\;\;,\;\;\;\;\;\;\;\;\;\;\;
U_0\;=\;c \;\langle \,0\,|\; U+U^\dagger\; |\,0\,\rangle\; \ne\; 0
\label{vev}
\end{equation}
that explicitly break  scale, respectively chiral symmetry.
In (\ref{Lchi}),
$b$ is related to the conventional bag constant $B$ by $B=b\chi_0^4/4$,
$c$ is a constant of mass dimension 3,  $m_q = \mbox{diag}(m_u,m_d,m_s)$
is the light quark mass matrix, and $m_0^2$ is an extra
U(1)-breaking mass term for the ninth pseudoscalar meson $\phi_0$.

Notice that the anomaly constraint (\ref{theta})
is modeled by the third and fourth term in (\ref{Lchi}) with the correspondence
\begin{equation}
\langle\,0\,|\,\frac{\beta(\alpha_s)}{4 \alpha_s} \,F_{\mu\nu}F^{\mu\nu}
\,|\,0\,\rangle \;=\;-\,b \;\chi_0^4
\label{gcond}
\end{equation}
and
\begin{equation}
\langle\,0\,|\; \bar q q \;|\,0\,\rangle \;=\;
c\,\left(\frac{\chi}{\chi_0}\right)^3 \;
\langle\,0\,|\,U + U^\dagger\,|\,0\,\rangle
\label{qcond}
\;.
\end{equation}
\medskip

\noindent {\bf 2.3 {\boldmath $\cal{L}[\psi,A,\chi]$}}
\medskip

The key ingredient in (\ref{S}) is the connection between the
scale and chiral symmetric,
short distance regime of colored fluctuations and the world of colorless
hadrons with broken symmetries. It is mediated by
the coupling between the fundamental quark and gluon degress and the
collective fields $\chi$ and $U$ through  coupling functions $g(\chi)$ and
$\xi(\chi)$,
\begin{equation}
\cal{L}[\psi,A,\chi] \;=\;
 \frac{\xi(\chi)}{4}\,\,F_{\mu\nu, a} F^{\mu\nu}_a
\;-\;  \overline{\psi}_i \,g(\chi)\, \psi_i
\label{LQCDchi}
\;.
\end{equation}
The coupling functions $\xi(\chi)$ and $g(\chi)$
are chosen in the spirit of Friedberg and Lee \cite{FL},
who formulated a {\it dual QCD vacuum picture}:
High momentum, short distance quark-gluon fluctuations (the perturbative
vacuum) are embedded
in a collective background field $\chi$ (the non-perturbative vacuum), in which
by
definition the low momentum, long range fluctuations are absorbed.
Confinement is thus associated with the colordielectric stucture of the QCD
vacuum.
This property is modeled by a colordielectric function
\begin{equation}
\kappa(\chi)\;=\; 1\;-\; \xi(\chi)
\label{kappa}
\end{equation}
that satisfies
\begin{equation}
\kappa(0)\;=\; 1
\;\;,\;\;\;\;\;\;\;\;\;\;\;\;\;\;
\kappa(\chi_0)\;=\; 0
\;,
\label{kappa1}
\end{equation}
thereby generating color charge confinement, because
a color electric charge creates a displacement $\vec D_a = \kappa \vec E_a$,
where $E_a^k = F_a^{0k}$, with energy $\frac{1}{2} \int d^3 r D_a^2/\kappa$
which is infinite for non-zero total charge if $\kappa$ falls off faster
than $1/\sqrt{r}$ for large distances $r$.
The particular form of $\kappa(\chi)$ is not crucial as long as the properties
(\ref{kappa1}) are satisfied \cite{wilets}.
A specific choice  is \cite{bickeboeller1}
\begin{equation}
\kappa(\chi)\;=\; 1\;+\; \left(\frac{\chi}{\chi_0}\right)^3 \;
\left(3\,\frac{\chi}{\chi_0} \;-\;4\right)\;\theta(\chi)
\label{kappa2}
\;,
\end{equation}
which has the further properties of
$\kappa^\prime(0)=\kappa^{\prime\prime}(0)=0$.
Other forms used in the literature are e.g. $\kappa(\chi)=|1
-(\chi/\chi_0)^n|^m$
\cite{bickeboeller2} (Friedberg and Lee originally proposed $n=m=1$).

Similarly, absolute confinement can be ensured also for quarks by coupling
the quark fields to the $\chi$-field through \cite{fai88}
\begin{equation}
g(\chi) \;=\; g_0 \;\left(\frac{1}{\kappa(\chi)}\;-\;1\right)
\;,
\label{gchi}
\end{equation}
which leads to an effective confinement potential with the masses of the quarks
inside
approximately equal to the current masses, but at $\chi=\chi_0$ it
generates an infinite asymptotic quark mass (the value of $g_0$ is
irrelevant in the present paper).
\bigskip

\noindent {\bf 2.4 Equations of motions}
\medskip

To summarize to this end,
the complete effective action (\ref{S}) is determined by
the Lagrangian
\begin{eqnarray}
\cal{L}_{eff} &=&
\cal{L}[\psi,A] \;+\; \cal{L}[\chi,U]  \;+\; \cal{L}[\psi,A,\chi]
\nonumber \\
&=& -\frac{1}{4}\,\kappa(\chi)\,F_{\mu\nu, a} F^{\mu\nu}_a
\;+\;  \overline{\psi}_i \left[\frac{}{}\,i \gamma_\mu \partial ^\mu \,
- g_s  \gamma_\mu A^\mu_a T_a \;
\;-\; g(\chi) \right]\, \psi_i
\nonumber \\
& &
+\; \frac{1}{2}\,(\partial_\mu \chi) ( \partial^\mu \chi )
\;+\; \frac{1}{4}\,Tr\left[\frac{}{}(\partial_\mu U)
( \partial^\mu U^\dagger ) \right]
\;\,-\;\, V(\chi,U)
\label{L}
\;,
\end{eqnarray}
plus  the terms $\cal{L}_{gauge}$ and $\cal{L}_{ghost}$ of (\ref{LQCD}).
The  potential $V$ is given by
\begin{eqnarray}
V(\chi, U) &=&
b \;\left[ \frac{1}{4} \,\chi_0^4 \;+\;
\chi^4 \,\ln\left(\frac{\chi}{e^{1/4} \chi_0}\right)\right]
\nonumber \\
& & +\;
 c\; Tr\left[\frac{}{} \hat m_q (U + U^\dagger) \right]
 \; \left(\frac{\chi}{\chi_0}\right)^3
\;+\;
\frac{1}{2}\; m_0^2 \; \phi_0^2
\;\left(\frac{\chi}{\chi_0}\right)^4\;
\;,
\label{V}
\end{eqnarray}
which has its minimum when $\chi=\chi_0=\langle 0|\chi|0\rangle$
and equals the vacuum pressure (bag constant) $B=b\chi_0^4/4$ at $\chi=0$.
Typical forms of $V(\chi,U)$ for different values of $B$ and $m_q$ are depicted
in Fig. 1.

The effective field theory defined by (\ref{S}) and (\ref{L})
represents a description of the duality of
partonic and hadronic degrees of freedom by coupling the
high energy QCD phase with unconfined gluon and quark degrees of freedom to a
low energy QCD  phase with confinement and broken chiral
symmetry which contains a gluon condensate (\ref{gcond}) and a
quark-antiquark condensate (\ref{qcond}).
Small oscillations about the minimum of the potential $V(\chi,U)$
are to be interpreted
as physical hadronic states that emerge after symmetry breaking. They
include \cite{ellis90}:
(i) glueballs and hybrids as quantum fluctuations in the gluon condensate
$\chi_0$,
(ii) pseudoscalar mesons as excitations of the quark condensate $U_0$,
(iii) the pseudoscalar flavor singlet meson $\phi_0$, and
(iv) baryons as non-topological solitons \cite{skyrme62}.
\smallskip

The field equations which derive from (\ref{S}) and (\ref{L}) are:
\begin{eqnarray}
& &
\left[ \gamma_\mu \left( i \partial^\mu \,-\, g_s A^\mu_a T_a \right) \;-\;
g(\chi) \right] \;\psi_i \;=\;0
\label{eom1}
\\
& &
\partial_\mu \left[ \kappa(\chi) \, F^{\mu\nu}_a\right]\;=\;
-g_s \,\kappa(\chi) \, f_{abc} \, A_{\mu, b} F^{\mu\nu}_c
\;+\;
g_s \,\overline{\psi}_i \,\gamma^\nu T_a \,\psi_i
\label{eom2}
\\
& &
\partial_\mu \partial^\mu \chi \;+\;
\frac{\partial V(\chi,U)}{\partial \chi}\;+\; \frac{\partial
\kappa(\chi)}{\partial \chi}\,
F_{\mu\nu, a} F^{\mu\nu}_a \;+\;
g(\chi) \,\overline{\psi}_i \psi_i \;=\;0
\label{eom3}
\\
& &
\partial_\mu \partial^\mu U \;+\;
\frac{\partial V(\chi,U)}{\partial U}\;=\;0
\label{eom4}
\;.
\end{eqnarray}
Notice that the $U$-field does not couple directly to the
quark and gluon fields. Per construction \cite{ellis90}, the dynamics of
the quark condensate field $U$ is solely driven by the
gluon condensate field $\chi$. It is important to realize that
the interplay between the $\chi$-field and the quark and gluon fields,
$\psi$ and $A$, is the crucial element of this approach.
\bigskip

\noindent {\bf 2.5 Comments}
\medskip

The following remarks concerning the effective
Lagrangian (\ref{L}) are important:
\begin{itemize}
\item[a)]
At short distances or high momentum transfers
the exact QCD Lagrangian (\ref{LQCD}) is recovered, since
$\chi=U=0$ and $\kappa(\chi)=1$ (i.e. $\xi(\chi)=0$) and $g(\chi)=0$,
whereas the long distance QCD properties
emerge as $\chi/\chi_0 \rightarrow 1$ and  $U/U_0 \rightarrow 1$ \cite{ellis90}
and no colored quanta survive.
The transition from one set of degrees of freedom ($\psi, A$) to
the other ($\chi, U$) corresponds to consecutively integrating
out all colored quantum fluctuations and absorbing them
effectively in the collective color singlet fields.

\item[b)]
The problem of double counting degrees of freedom has to be carefully
inspected. Although it does not arise in one-loop calculations (to which I will
restrict here), processes with e.g. two-gluon exchange
could also be contained in the exchange of a color singlet $\chi$-quantum.
A minimal possibility to avoid this problem is a rigid separation
of high and low momentum modes, by introducing a characteristic
scale $Q_0$: above $Q_0$ the physics is described in terms of quark gluon
degrees and below $Q_0$ the dynamics is governed by the
collective degrees of freedom \cite{ellwanger94}.

\item[c)]
$\cal{L}[\chi,U]$ for the composite
fields embodies the correct
QCD scaling and chiral properties and accounts for the important anomaly
(\ref{theta})
of the physical energy-momentum tensor of QCD.
The coupling between quarks and gluons to the composite field $\chi$
in $\cal{L}[\psi,A,\chi]$
can be interpreted in analogy to a thermodynamic system in equilibrium with a
heat bath,
with a net flow of energy between the system and the heat bath environment
such that the {\it bare} energy of the system is not conserved.
However the {\it free} energy of the system, here high momentum quarks and
gluons,
is constant \cite{bass94}. It corresponds to the conserved  energy momentum
tensor $\theta_{\mu\nu}$ with its non-zero trace (\ref{theta}).

\item[d)]
There is no need for explicit renormalization of
$\Delta \cal {L}=\cal{L}[\chi,U]+\cal{L}[\psi,A,\chi]$.
The composite fields  $\chi$ and $U$ are already interpreted as effective
degrees of freedom with loop corrections implicitely included in $\Delta
\cal{L}$
and it would be double counting to add them again.
Moreover, in the present approach
the low energy domain of $\cal{L}[\chi,U]$
is per construction bounded from above by the onset of the high energy regime
described by $\cal{L}[\psi,A]$.
In correspondance to item b) the scale $Q_0$ that separates the two
domains, provides an "ultra-violet" cut-off for
$\cal{L}[\chi,U]$, and at the same time an infra-red cut-off for
$\cal{L}[\psi,A]$.
\end{itemize}
\smallskip

This effective field theory approach
offers a wide range of physical applications and
can be extended and refined in various directions, as discussed in Sec. 4.
The scope of the remainder of this paper is however concepted as a
exemplary demonstration of how of the dynamics of parton-hadron conversion
emerges within this framework.
\bigskip

\noindent {\bf 3. CONFINEMENT OF GLUONS IN A FRAGMENTING {\boldmath $q\bar q$}
SYSTEM}
\bigskip

The effective QCD field theory defined by (\ref{L}) is readily applicable
to describe the dynamic evolution from perturbative to non-perturbative
vacuum in high energy processes.
In accord with the symmetry breaking formalism of Sec. 2, the parton-hadron
transition can be visualized as the conversion of high momentum colored
quanta of the fundamental quark and gluon fields into color neutral
composite states that are described by the condensate fields $\chi$ and $U$
and their excitations.

In the following I shall
consider as an example the fragmentation of a $q\bar q$ jet system
with its emitted bremsstrahlung gluons and describe the evolution of the
system as it converts from the parton phase to the hadronic phase.
The process is illustrated in Fig. 2:
A time-like virtual photon in an $e^+e^-$ annihilation event
with large invariant mass $Q^2 \gg \Lambda^2$ is assumed to produce
a  $q \bar q$ pair which initiates a cascade of sequential gluon emissions.
(Here and in the following $\Lambda$ denotes the fundamental QCD scale).
The early stage is characterized by emission of "hot" gluons
far off mass shell in the perturbative vacuum. Subsequent gluon
branchings yield "cooler" gluons with successively
smaller virtualities, until they are within $Q_0^2$,
where $Q_0$ is of the order of $m_\chi\equiv 4b\chi_0^2\simeq 1$ GeV.
At this point condensation sets in, or loosely speaking, the "cool" gluons
are eaten up by the color neutral gluon condensate field $\chi$, the
particle excitations of which must then decay into physical hadrons
by means of some local interaction in the non-perturbative vacuum.
A similar picture holds for the $U$-field.

It is well known that the bulk of produced particles stems from
rather soft gluon emissions that are characterized by small values
$x$ of the fraction of the initial energy. Secondary
production of $q\bar q$ pairs is comparably rare on the perturbative
level.  It is therefore reasonable to neglect the quark degrees of
freedom and study the purely gluonic sector.
Furthermore it is convenient to work in  a physical (axial) gauge
for the gluon fields, $\eta \cdot A=0$, by choosing the gauge vector in
(\ref{LQCD})
as $\eta_\mu = n_\mu$ with $n$ being space-like and constant,
in which case the  ghost contribution vanishes.
Consequently,
the  equations of motion (\ref{eom1})-(\ref{eom4}) reduce to:
\begin{eqnarray}
& &
\partial_\mu \left[ \kappa(\chi) F^{\mu \nu}_a\right] \;=\;
- g_s \kappa(\chi) f_{abc} A_{\mu, b} F^{\mu\nu}_c
\label{e1}
\\
& &
\partial_\mu \partial^\mu \chi \;=\;
- \frac{\partial V(\chi)}{\partial \chi} \;-\;
\frac{\partial\kappa(\chi)}{\partial \chi}
\, F_{\mu\nu, a} F^{\mu\nu}_a
\label{e2}
\\
& &
\partial_\mu \partial^\mu U \;=\;
- \,4\,c \left(\frac{\chi}{\chi_0}\right)^3\;
Tr [m_q]
\label{e3}
\;.
\end{eqnarray}
The solution of these equations is  a still formidable task, because not
only the gluon fields but also $\chi$ and $U$ are quantum operators.
To make progress, I will now proceed by
(i) treating the quantum gluon fields perturbatively, and (ii)
employing the mean field approximation for the composite
fields $\chi$ and $U$.
Representing
\begin{equation}
\chi(r)\;=\;  \bar \chi(r) \;+\; \hat\chi(r)
\label{mfa}
\;,
\end{equation}
where $\bar\chi$ is a c-number and $\hat\chi$ a quantum operator
(similarly for $U$),
the mean field approximation is obtained by neglecting the
quantum fluctuations $\hat \chi$ and keeping only $\bar \chi$.
Thus, the approximations (i) and (ii) correpond to the semiclassical
limit in which gluonic quantum fluctuations interact with
a classical mean field.
To a good approximation this should provide a reasonable
description: first, because renormalization group improved
QCD perturbation theory allows for an accurate description of
the evolution of the gluon field \cite{mueller81} at short  distances
where $\chi= 0$, and second,
because the dynamics of the system around $\chi=\bar \chi$ is
governed by a large number of virtual excitations,
corresponding to coherent modes of field quanta, so that
a quasiclassical mean field description should be applicable
in the low energy regime \cite{FL}.
\bigskip

\noindent {\bf 3.1 Evolution of the gluon distribution in the presence of the
collective field {\boldmath $\chi$}}
\medskip

As I will show now, the equations of motion (\ref{e1})-(\ref{e3}) simplify
to a perturbative evolution equation for the gluon distribution which is
coupled
to the equation for the mean field $\bar \chi$.
The key problem is the first equation, since the gluon fields $A^\mu$, or
equivalently $F^{\mu\nu}$, drive the
dynamics of the $\chi$-field which in turn feeds back via $\kappa(\chi)$.
As mentioned before, the $U$-field does not couple directly to the gluon field.
The procedure in the following is therefore to 'solve' eq. (\ref{e1})
as a function of $\kappa$ and then to insert the solution into (\ref{e2}),
so that one is left (aside from the simple third equation) with a single
equation for $\chi$, which however is non-linear.
\smallskip

Solving the equation of motion (\ref{e1}) for the gluon field is equivalent to
the calculation of the complete Greens function with an arbitrary number
of gluons.
Instead, I will restrict to evaluate the 2-point Greens function
only (Fig. 3), i.e.  the full gluon propagator which includes both the
one-loop order
gluon self-interaction through real and virtual emission and absorption,
and the effective interaction with the confining background field $\chi$.
In the framework of "jet calculus" \cite{konishi79}, this gluon propagator
denoted as $D_g(x,k^2;x_0,Q^2)$,
describes how a gluon, produced with an invariant mass $Q^2$,
 evolves in the variable $x$ (momentum or energy fraction) and the virtuality
$k^2$ (or transverse momentum $k_\perp^2$) through these interactions.
To  one-loop order, it is obtained by calculating the
corresponding cut diagrams. In the present case, one has
in addition to the usual gluon branching and fusion processes,
$g\rightarrow gg$ and $gg\rightarrow g$,
contributions from energy transfer and two gluon annihilation processes
$g\rightarrow g \chi$ and $gg \rightarrow \chi$, respectively
\footnote{In the present case of $q\bar q$ jet evolution, the contribution of
perturbative 2-gluon fusion
processes $gg\rightarrow g$ for $k^2 \gg\Lambda^2$ is very small
\cite{muellqiu}, but the non-perturbative gluon recombination
$gg\rightarrow \chi$ in the range
$Q_0^2\,\lower3pt\hbox{$\buildrel > \over\sim$}\, k^2 \ge \Lambda^2$
is of essential importance in order to achieve complete confinement.}.
This is illustrated in Fig. 3.

To write down the determining equation for the
gluon propagator $D_g$, it is convenient
employ lightcone variables, defined by the identification of components of
four-vectors as
\begin{equation}
k^\mu \;=\; (k^+, k^-, \vec k_\perp )
\;\;;\;\;\;\;\;\;
k^\pm \;=\; \frac{1}{\sqrt{2}} \,(k^0 \pm k^3 )
\;\;;\;\;\;\;\;\;
k^\perp \;=\; (k^1, k^2)
\;\;.
\label{lc1}
\end{equation}
The $k^+$ component of a particle's momentum,
the {\it light cone momentum}, is always positive definite,
$k^+ > 0$, and the {\it light cone energy}, $k^- = (k_\perp^2 + m^2)/(2k^+)$
is also positive.
Furthermore, the {\it light cone time} $ r^+ = (t+z)/\sqrt{2}$ is conjugate
to $k^-$ and the {\it light cone coordinate} $r^-$ is conjugate to $k^+$.
The invariant momentum space element is
\begin{equation}
\frac{d^3 k}{(2 \pi)^3 2 E} \;=\; \frac{d^4 k}{(2\pi)^4} \,\delta^+(k^2 - m^2)
\;
=\; \frac{d k^+ \,d^2 k_\perp}{16 \pi^3 \,  k^+}
\;.
\label{lc3}
\end{equation}
Choosing the {\it light cone gauge} for the gluon fields,
$\eta \cdot A = A^+ =- A^- = 0$,
results in well known simplifications in the perturbative
analysis of light cone dominanted processes and has the advantage that there
are neither negative norm gluon states nor ghost states present
\cite{brodsky91}.
As a consequence only the transverse components $A^i_\perp$ ($i=1,2$) are
dynamical field variables, since $A^+$ is identically zero and $A^-$ is
determined at any 'time' $r^+$ by $A^1_\perp$ and $A^2_\perp$.
The particular choice $\eta = (p_q + p_{\bar q})/2$ has the advantage that
interference terms do not contribute \cite{pQCD2} to leading log accuracy
(they are suppressed $\propto 1/k^2$).
Therefore, in the leading log approximation (LLA)
\cite{lla1,dok80,mueller81,lla2},
it is enough to realize that for every
choice of $b$ (Fig. 3 top), one can group the other gluons in a unique way to
groups forming dressed rungs of a ladder (Fig. 3 middle) whose discontinuity is
taken (Fig. 3 bottom).

Introducing the variable $x=k^+/P^+$ (the light cone fraction), and
parametrizing the momenta of initial quark and antiquark as $P\equiv p_q +
p_{\bar q}$ with
\begin{equation}
P^+ \;=\; Q
\;\;,\;\;\;\;\;
P^- \;=\; \frac{4m_q^2}{2 P^+}
\;\;,\;\;\;\;\;
\vec P_\perp \;=\; \vec 0
\;\;,
\label{PQQ}
\end{equation}
i.e. $P^2 = Q^2$,
where as before $Q$ denotes the invariant mass of the time-like photon
that creates the pair,
the determining equation for the gluon propagator
$D_g(x,x_0;k^2,Q^2)$ can now be represented in the form
(c.f. Appendix A):
\begin{eqnarray}
D_g(x,k^2;x_0,Q^2) &=&
x_0 \delta(x-x_0)\,\delta(k^2 - Q^2)\;F_g(Q^2,Q_0^2)
\nonumber \\
&+&
 F_g(Q^2,k^2) \int_{Q_0^2}^{k^2}\frac{d k^{'\,2}}{k^{'\,2}}
\int_x^1 \frac{dx'}{x'} w(x',x,k^2)\;D_g(x',k^{'\,2};x_0,Q^2)\;
F_g(k^{'\,2},Q_0^2)
\nonumber \\
&&
\label{ee0}
\;.
\end{eqnarray}
This equation has a simple physical significance:
The first term is the inclusive sum of virtual
emissions and reabsorptions, and therefore does not change the number of
gluons in the gluonic wavefunction of the fragmenting $Q\bar Q$ pair,
whereas the second term describes the change of the gluon distribution
as a result of real decay or fusion processes.
The {\it Sudakov formfactor}
\begin{equation}
F_g(Q^2,k^2) \;=\;\exp \left[-\, \int_{k^2}^{Q^2}\frac{dk^{'\,2}}{k^{'\,2}}
w_g(k^{'\,2})\right]
\,
\label{Fg1}
\end{equation}
is the probability
a gluon propagates like a  bare, non-interacting particle
while degrading its virtuality from $Q^2$ to $k^2$. As the gap between $Q^2$
and $k^2$ grows, such a fluctuation becomes increasingly unlikely.
The {\it total interaction probability}
\begin{equation}
w_g(k^2)\;=\;\int_0^1dx \int_x^1 \frac{dx'}{x'} \;w(x',x,k^2)
\end{equation}
is the integral over the inclusive probability for all possible
gluon interaction processes $i$,
\begin{equation}
w(x',x,k^2)\;=\;\sum_{proceses\;i} w_{(i)}(x',x,k^2)
\;.
\end{equation}
The normalization is such that
\begin{equation}
1\;=\;F_g(Q^2,Q_0^2) \;+\;
F_g(Q^2,k^2) \;\int_{Q_0^2}^{Q^2}\frac{d k^{'\,2}}{k^{'\,2}}\,
w_g(k^{'\,2})\; F_g(k^{'\,2},Q_0^2)
\label{eenorm}
\end{equation}
in accord with unitarity (probability) conservation.
Multiplying (\ref{ee0}) by $F_g^{-1}$, differentiating, and accounting
for (\ref{eenorm}), yields:
\begin{equation}
k^2
\frac{\partial}{\partial k^2} \,D_g(x,k^2;x_0,Q^2) \;=\;
\int_x^1 \frac{dx'}{x'} w(x',x,k^2)\; D_g(x',k^2;x_0,Q^2)\;
-\; w_g(k^2)\;D_g(x,k^2;x_0,Q^2)
\label{ee1}
\end{equation}

As summarized in the Appendices, one obtains for the individual
interaction probabilities $w_{(i)}$ to one loop order
(with the assignment
$x_1\rightarrow x_2, (x_1-x_2)$ for branchings and
$x_1, (x_2-x_1)\rightarrow x_2$ for fusions)
\begin{eqnarray}
w_{g\rightarrow gg}(x_1,x_2,k^2) &=&
\frac{\alpha_s(k^2)}{2\pi}\;\gamma_{g\rightarrow
gg}\left(\frac{x_2}{x_1}\right)
\nonumber \\
w_{g\rightarrow gg}(x_1,x_2,k^2) &=&
\frac{\alpha_s(k^2)}{2\pi}\;
\left[ \frac{8\pi^2 \,c_{gg\rightarrow g}}{k^2 \Lambda^2}\;
\; \frac{x_1(x_2-x_1)}{x_2^2}\right]
\;\gamma_{g\rightarrow gg}\left(\frac{x_1}{x_2}\right)
\nonumber \\
w_{g\rightarrow g \chi}(x_1,x_2,k^2) &=&\;
\frac{\lambda_\chi(k^2)}{2\pi}\;
\gamma_{g\rightarrow g\chi}\left(\frac{x_2}{x_1}\right)
\nonumber \\
w_{gg \rightarrow \chi}(x_1,x_2,k^2) &=&
\frac{\lambda_\chi(k^2)}{2\pi}\;
\left[ \frac{8\pi^2 \,c_{gg\rightarrow \chi}}{k^2 \Lambda^2}\;
\; \frac{x_1(x_2-x_1)}{x_2^2}\right]
\;\gamma_{\chi\rightarrow gg}\left(\frac{x_1}{x_2}\right)
\label{w}
\;,
\end{eqnarray}
where $c_{gg\rightarrow g} = c_{gg\rightarrow \chi} = 1/8$, and
\begin{eqnarray}
\gamma_{g \rightarrow g g} (z) &=&
2\,C_A\;\left( z ( 1 - z ) + \frac{z}{1-z} + \frac{1-z}{z} \right)
\nonumber
\\
\gamma_{g \rightarrow g \chi} (z) &=&
\frac{1}{4} \left(\frac{ 1\;+\;z^2}{1\;-\;z}\right)
\nonumber
\\
\gamma_{\chi \rightarrow g g} (z) &=&
8 \left(\frac{}{} z^2 \, - \, z \, +\, \frac{1}{2}\right)
\label{gamma}
\;.
\end{eqnarray}
Here $C_A = N_c = 3$, and $z$ is the fraction of $x$-values of daughter to
mother gluons.
In (\ref{w}),
\begin{equation}
\alpha_s (k^2)\;=\;
\left[ b \; \ln \left(\frac{k^2}{\Lambda^2}\right)\right]^{-1}
\;\;,\;\;\;\;\;\;\;\;\;\;
b\;=\;\frac{11 N_c - 2 N_f}{12\pi}
\;\;\;,
\label{alpha}
\end{equation}
is the one loop order QCD running coupling (in the present case $N_f =0$), and
\begin{equation}
\lambda_\chi (k^2)\;=\;
\frac{\tilde{\xi}_\chi^2(k^2)}{4\pi}
\label{lambda}
\end{equation}
denotes the coupling to the $\chi$-field in momentum space,
with $\tilde{\xi}_\chi$ denoting the Fourier transform of $\xi(\chi)$
in (\ref{LQCDchi}).

Eq. (\ref{ee0}) for the propagator $D_g(x,k^2;x_0,Q^2)$
corresponds the evolution equation for the gluon
distribution $g(x,k_\perp^2)$, which is defined \cite{collins82} as
average number of gluons at 'light cone time' $r^+ =0$ in
the multi-gluon state\footnote{It is convenient to visualize the
initial $q\bar q$ pair (\ref{PQQ}) as a single incoming
'gluon' with momentum $P$, i.e. with $x_0=1$ and invariant mass $Q$ (c.f Fig.
3).}
$|P\rangle$
with light cone fractions $x\equiv k^+/P^+$ in a range $dx$ and
transverse momenta in a range $d^2 k_\perp$:
\begin{eqnarray}
x\,g(x, k_\perp^2)
&=&
\frac{1}{P^+}
\int \frac{d r^- d^2 r_\perp}{(2\pi)^3} \;
e^{-i (k^+ r^- -\vec k_\perp \cdot \vec r_\perp)} \;
\langle P |
\,F_a^{+ \mu}(0,r^-,\vec r_\perp)\left. F_{a,\,\mu}\right. ^+(0,0,\vec
0_\perp)\,
|P \rangle
\;.
\label{gdef}
\end{eqnarray}
Thus, as is evident from Fig. 3, the probability for finding a gluon with $x$
and $k_\perp^2$ is
given by the identification
\begin{equation}
g(x,k_\perp^2)\;=\; \left. \frac{}{}D_g(x,k^2; 1, Q^2)\right|_{k^2=k_\perp^2}
\;.
\end{equation}
On account of the explicit  expressions (\ref{w})-(\ref{lambda}),
and taking as evolution variable the gluon transverse momentum
$k_\perp^2$ rather than the invariant mass $k^2$ \cite{basetto83},
one finally arrives at the master equation for the
evolution of the gluon distribution:
\begin{eqnarray}
k_\perp^2 \frac{\partial}{\partial k_\perp^2} \,g(x,k_\perp^2)
&=&
+\;\frac{\alpha_s(k_\perp^2)}{2 \pi} \;
\int_0^1 dz \,\left[\frac{1}{z} \, g\left(\frac{x}{z},k_\perp^2\right) \;-\;
\frac{1}{2} \; \,g(x,k_\perp^2)\right]\; \gamma_{g\rightarrow gg}\left(z\right)
\;\theta\left(k_\perp^2 - Q_0^2\right)
\nonumber \\
& & -\;
\frac{\alpha_s(k_\perp^2)}{2 \pi} \;
\left(\frac{\Lambda^2}{k_\perp^2}\right)\;
8 \pi^2 \, c_{gg\rightarrow g}
\;\int_0^1 dz \, \left[
g^{(2)}\left(x, \frac{1-z}{z}x, k_\perp^2\right)
\right.
\nonumber \\
& & \;\;\;\;\;\;\;\;\;\;\;\;\;\;\;\;\;\;\;\;\;\;\;\;\;
\left.
\;-\;
\frac{1}{2}
g^{(2)}\left(z x,(1-z)x, k_\perp^2\right)
\right]\;
\gamma_{g\rightarrow gg}\left(z\right)
\;\theta\left(k_\perp^2 - Q_0^2\right)
\nonumber \\
& & +\;
\frac{\lambda_\chi(k_\perp^2)}{2 \pi} \;
\int_0^1 dz \,\left[\frac{1}{z} \, g\left(\frac{x}{z},k_\perp^2\right)\;-\;
g(x,k_\perp^2)\right]\; \gamma_{g\rightarrow g\chi}\left(z\right)
\label{ee2} \\
& & -\;
\frac{\lambda_\chi(k_\perp^2)}{2 \pi} \;
\left(\frac{\Lambda^2}{k_\perp^2}\right)\;
8 \pi^2 \, c_{gg\rightarrow \chi}
\;\int_0^1 dz \,
g^{(2)}\left(x, \frac{1-z}{z}x, k_\perp^2\right)
\;\gamma_{\chi\rightarrow gg}\left(z\right)
\;.
\nonumber
\end{eqnarray}
This equation reflects the probabilistic
parton cascade interpretation of the LLA \cite{durand85,collins87,ms17},
in which the change of the gluon distribution on the left hand side
is governed by the balance of gain (+) and loss ($-$) terms on the right hand
side.
Notice that (\ref{ee2}) is free of infrared divergences, because the
singularities in (\ref{gamma}) at $z=0$ and $z=1$ cancel between
gain and loss terms.  A diagrammatic illustration of these
gain and loss terms is shown in Fig. 4.
Also notice that
the gluon fusion terms (second and fourth term) involve the
2-gluon distribution $g^{(2)}(x_1,x_2,k_\perp^2)$,
the presence of which causes the evolution equation
to be non-linear.

By means of a Mellin transformation the
multiple convolutions of $z$-integrals inherent in
the iteration of  eq. (\ref{ee2}) can
be converted into products of independent successive
interaction probabilities.
Let me define the gluon distribution in the {\it moment representation}
as
\begin{equation}
g(\omega,k_\perp^2)\;:=\; \int_0^1 dx \, x^\omega\; g(x,k_\perp^2)
\;\,=\;\,
\int_0^\infty dy \, e^{-\omega\,y} \left[ x g(x,k_\perp^2)\right]
\;,
\label{ome1}
\end{equation}
where $y=\ln(1/x)$, in which the variable $\omega$
is conjugate to $\ln(1/x)$, implying that the low $x$ behaviour
is determined by small values of $\omega$.
Analogously, the 2-gluon distribution is represented as
\begin{equation}
g^{(2)}(\omega,z,k_\perp^2)\;:=\; \int_0^1 dx \, x^\omega\;
g^{(2)}\left(x, \frac{1-z}{z}x, k_\perp^2\right)
\;,
\label{ome2}
\end{equation}
which carries an additional $z$-dependence.
In general $g^{(2)}$ is a complicated correlation function not available in
analytical form, except for certain special cases \cite{konishi79,basetto80}.
It is therefore
inevitable on an analytical level to assume a phenomenological form of
$g^{(2)}$  that
allows to convert eq. (\ref{ee2}) into a  tractable linear form.
This can be achieved with the
following ansatz of product form \cite{muellqiu,close89}
\begin{equation}
g^{(2)}\left(x, \frac{1-z}{z}x, k_\perp^2\right)
\;=\;
\rho(k_\perp^2)\;
g\left(x,k_\perp^2\right)\;g\left(\frac{1-z}{z}x, k_\perp^2\right)
\;,
\label{g2x}
\end{equation}
where $\rho(k_\perp^2)$ is a parameter that characterizes the magnitude
of the probability for finding two gluons at the same point in phase-space,
depending on their typical transverse size $r_\perp\sim 1/k_\perp$.
Since the 2-gluon correlation must become substantial when the number of gluons
per unit area
$n_g/(\pi R_\perp^2)$ becomes so large that the gluons overlap spatially
($R_\perp \approx 0.5-1$ fm),
one expects that $\rho = O(1)$ when
$k_\perp^2\,\approx\, Q_0^2$ ($Q_0= 1$ GeV), and monotonically increasing
as $k_\perp^2 \rightarrow \Lambda^2$.
Using (\ref{g2x}) in (\ref{ome2}), the 2-gluon distribution in
the moment representation can be approximated in the soft limit ($z \ll 1$)
as
\begin{equation}
g^{(2)}(\omega,z,k_\perp^2)\;\approx\; \rho(k_\perp^2)
\;\frac{z^\omega}{\omega}\;
g(\omega,k_\perp^2)
\;.
\label{g2ome}
\end{equation}

In the moment representation the evolution equation (\ref{ee2})
now becomes an linear algebraic equation for the Mellin transformed
gluon distribution:
\begin{equation}
k_\perp^2 \frac{\partial}{\partial k_\perp^2} \, g(\omega,k_\perp^2)
\;=\; \gamma(\omega,k_\perp^2) \; \, g(\omega,k_\perp^2)
\;,
\label{ee5}
\end{equation}
where $\gamma(\omega,k_\perp^2)$ plays the role of
a generalized {\it anomalous dimension},
\begin{eqnarray}
\gamma(\omega,k_\perp^2)&=&
\frac{\alpha_s(k_\perp^2)}{2 \pi} \; \theta(k_\perp^2 -Q_0^2)\;
\left[ A_{g \rightarrow gg}(\omega) \;-\;
\left(\frac{\Lambda^2}{k_\perp^2}\right)
\,A_{gg\rightarrow g}(\omega)\right]
\nonumber \\
& &
+\;\frac{\lambda_\chi(k_\perp^2)}{2 \pi} \;
\left[ A_{g \rightarrow g\chi}(\omega) \;-\;
\left(\frac{\Lambda^2}{k_\perp^2}\right)
\,A_{gg\rightarrow \chi}(\omega)\right]
\nonumber \\
&\equiv &
\gamma_{QCD}(\omega,k_\perp^2)
\;+\;
\gamma_\chi(\omega,k_\perp^2)
\;.
\label{andim}
\end{eqnarray}
The functions $A(\omega)$ are given by
\begin{eqnarray}
A_{g\rightarrow gg}(\omega) &=&
2\;C_A \;\left[\frac{11}{12} + \frac{1}{\omega} -
\frac{2}{\omega+1} + \frac{1}{\omega +2}
-\frac{1}{\omega+3}  - S(\omega)\right]
\nonumber \\
A_{gg\rightarrow g}(\omega) &=&
\frac{\pi^2\;\rho}{\omega} \; A_{g\rightarrow gg}(\omega)
\nonumber \\
A_{g\rightarrow g\chi}(\omega) &=&
\frac{1}{4}\;\left[\frac{3}{2} - \frac{1}{\omega+1} - \frac{1}{\omega +2}
 - 2\, S(\omega)\right]
\nonumber \\
A_{gg\rightarrow \chi}(\omega) &=&
\frac{\pi^2 \;\rho}{2\,\omega} \;
\left[\frac{1}{\omega+1} - \frac{2}{\omega +2} - \frac{2}{\omega +3}\right]
\;.
\label{A}
\end{eqnarray}
where
$S(\omega) = \psi(\omega+1)- \psi(1)$
with the Digamma function  $\psi(z) = d[\ln \Gamma(z)]/dz$ and
$- \psi(1) =  \gamma_E= 0.5772$  the Euler constant.

The anomalous dimension $\gamma(\omega,k_\perp^2)$, eq. (\ref{andim}),
 reduces at $k_\perp^2 \gg \Lambda^2$ to $\gamma_{QCD}$,
the  QCD anomalous dimension in the LLA \cite{basetto83},
since in this kinematic region $\kappa(\chi) = 1$ and therefore $\lambda_\chi =
0$.
However, at $k_\perp^2\simeq Q_0^2$, $\lambda_\chi(k_\perp^2)$
becomes non-zero, so that the evolution of the
gluon distribution receives modifications due to the coupling of gluons to the
$\chi$-field.  In the region
$Q_0^2 > k_\perp^2 \ge \Lambda^2$, the perturbative QCD contributions
$\propto \alpha_s$ vanish per construction, so that
the gluons solely interact with the $\chi$-field, the coupling
to which increases, because $\kappa \rightarrow 0$, i.e.  $\lambda_\chi
\rightarrow 1$.
This behaviour is evident in Fig. 5, which  shows $\gamma(\omega,k_\perp^2)$
versus $\omega$ for different values of $k_\perp^2$.

The formal solution of eq. (\ref{ee5}) is
\begin{equation}
g(\omega,k_\perp^2)\;=\;
\exp\left\{\int_{k_\perp^2}^{Q^2} \frac{d k_\perp ^{\prime\,2}}{k_\perp
^{\prime\,2}}
 \; \gamma(\omega ,k_\perp^{\prime\,2})\right\}
\label{sol1}
\end{equation}
from which the $x$ distribution can be reconstructed by considering
the inverse Mellin transform
\begin{equation}
x \,g(x,k_\perp^2) \;=\; \frac{1}{2\pi i}\;\int_C d\omega \,x^{-\omega}
g(\omega,k_\perp^2)
\label{sol2}
\;,
\end{equation}
where $\omega$ is now a complex variable and the contour of the integration $C$
runs paralell to the imaginary axis.
For the full anomalous dimension (\ref{andim}) this inversion must be done
numerically \cite{rkellis94}.
\bigskip

\noindent {\bf 3.2 Analytical estimates for {\boldmath $x$}-spectra and gluon
multiplicity}
\medskip

To exhibit the main features of the evolution of the gluon distribution
in the presence of the $\chi$-field, it is instructive to make some analytic
estimates. Of particular interest is the low $x$-region, because the
soft, small $x$  gluons are most preferably radiated, but at the same
time take away only a very small portion of the total energy.
For simplicity I will divide the kinematic range of the $k_\perp^2$-evolution
into two distinct domains as indicated in  Fig. 2:
\begin{itemize}
\item[{\bf (i)}]
$Q^2 \ge k_\perp^2 > Q_0^2$ and $k_\perp^2/\Lambda^2 \gg 1$:
In this region $\lambda_\chi\simeq 0$
and $\partial \lambda_\chi /\partial \ln k_\perp^2\simeq 0$,
so that (\ref{andim}) reduces to
\begin{equation}
\gamma(\omega,k_\perp^2)\;=\;
\frac{\alpha_s(k_\perp^2)}{2 \pi} \; A_{g \rightarrow gg}(\omega)
\label{ee6a}
\;.
\end{equation}
\item[(ii)]
$Q_0^2 \ge k_\perp^2 \ge \Lambda^2$ and $k_\perp^2/\Lambda^2 \rightarrow 1$:
Here $\lambda_\chi\rightarrow (4\pi)^{-1}$
and $\partial \lambda_\chi /\partial \ln k_\perp^2 > 0$,
in which case (\ref{andim}) becomes
\begin{equation}
\gamma(\omega,k_\perp^2)\;=\;
\;\frac{\lambda_\chi(k_\perp^2)}{2 \pi} \;
\left[ A_{g \rightarrow g\chi}(\omega) \;-\;
\left(\frac{\Lambda^2}{k_\perp^2}\right)
\,A_{gg\rightarrow \chi}(\omega)\right]
\;.
\label{ee6b}
\end{equation}
\end{itemize}

As stated before, the low $x$ region corresponds to the limit
$\omega \rightarrow 0$, so that an expansion of $\gamma(\omega,k_\perp^2)$
around $\omega=0$ gives the dominant contributions at small $x$.
Up to to order $\omega$ one has in the region (i)
\begin{equation}
k_\perp^2 \frac{\partial}{\partial k_\perp^2} \, \ln g(\omega,k_\perp^2)
\;=\; \frac{\alpha_s(k_\perp^2)}{2\pi} \;
\left[2 C_A\,\left(\frac{1}{\omega} - \frac{11}{12}\right)\right]
\;\equiv\; \gamma^{(i)}(\omega,k_\perp^2)
\;,
\label{ee7a}
\end{equation}
whereas in the region (ii) one gets in the same order of approximation
\begin{equation}
k_\perp^2 \frac{\partial}{\partial k_\perp^2} \, \ln g(\omega,k_\perp^2)
\;=\; -\;\frac{\lambda_\chi(k_\perp^2)}{2\pi} \;
\rho\,\frac{\Lambda^2}{k_\perp^2}\;
\left[\frac{8 \pi^2}{3\omega} \right]
\;\equiv\; \gamma^{(ii)}(\omega,k_\perp^2)
\;.
\label{ee7b}
\end{equation}
The accuracy of the expressions $\gamma^{(i)}$ and $\gamma^{(ii)}$
in the relevant range of $\omega$ is exhibited in Fig. 6, where the exact
expression $\gamma(\omega,k_\perp^2)$ is compared to the small-$\omega$
expansions (\ref{ee7a}) and (\ref{ee7b}) at large and small $k_\perp^2$.
Evidently the approximation is rather good for $\omega < 2$.

Eqs. (\ref{sol1}) and (\ref{sol2}) can now be solved analytically with the
saddle point method.
The $k_\perp^2$-dependence of $\alpha_s$ is given in (\ref{alpha}) and
for $\lambda_\chi = \tilde \xi_\chi^2/(4\pi)$, eq. (\ref{lambda}), I will use
here the form
\begin{equation}
\lambda_\chi(k_\perp^2)\;=\;
\frac{\theta(Q_0^2 -
k_\perp^2)}{4\pi}\;\frac{\ln(Q_0^2/k_\perp^2)}{\ln(Q_0^2/\Lambda^2)}
\label{lambda2}
\;,
\end{equation}
which has the required properties
that $\lambda_\chi = 0$ ($\tilde \xi_\chi =0$) for $k_\perp^2 \ge Q_0^2$
and $\lambda_\chi \rightarrow  (4\pi)^{-1}$ ($\tilde \xi_\chi \rightarrow 1$)
for $k_\perp^2 \rightarrow \Lambda^2$
(as before $\tilde{\xi}_\chi$ denotes the Fourier transform of $\xi(\chi)$ in
eq. (\ref{LQCDchi})).
Introducing the variables $\hat t$ for the region (i) and $\hat u$ for the
region (ii),
\begin{eqnarray}
\hat t (k_\perp^2) &=&\int_{k_\perp^2}^{Q^2} \frac{d
k_\perp^{'\,2}}{k_\perp^{'\,2}}
\,\frac{\alpha_s(k_\perp^{'\,2})}{2\pi}
\;=\;\frac{1}{2\pi
\,b}\;\ln\left[\frac{\ln(Q^2/\Lambda^2)}{\ln(k_\perp^2/\Lambda^2)}\right]
\nonumber
\\
\hat u (k_\perp^2) &=&\int_{k_\perp^2}^{Q_0^2} \frac{d
k_\perp^{'\,2}}{k_\perp^{'\,2}}
\,\left(\frac{\lambda_\chi(k_\perp^{'\,2})}{2\pi}
\;\rho\,\frac{\Lambda^2}{k_\perp^2}\right)
\;=\;
\frac{\rho}{8\pi^2}\;
\frac{1\,-\,(Q_0^2/k_\perp^2) \left(\frac{}{}1 - \ln (Q_0^2/k_\perp^2)\right)}
{(Q_0^2/\Lambda^2)\ln (Q_0^2/\Lambda^2)}
\;,
\label{u}
\end{eqnarray}
the combination of (\ref{sol1}) and (\ref{sol2}) yields
for the kinematic domains (i), respectively (ii):
\begin{eqnarray}
x g(x,\hat t)&=&\frac{1}{2\pi i}\,\int_C d\omega
\;\exp\left[ \omega \,y \;+\; \nu^{(i)}(\omega)\,\hat t \right]\;\,g(\omega,0)
\;,\;\;\;\;\;
\nu^{(i)}(\omega)\;=\;
2 C_A\,\left(\frac{1}{\omega} - \frac{11}{12}\right)
\nonumber
\\
x g(x,\hat u)&=&\frac{1}{2\pi i}\,\int_C d\omega
\;\exp\left[ \omega \,y \;+\; \nu^{(ii)}(\omega)\,\hat u \right]\;\,g(\omega,0)
\;,\;\;\;\;\;
\nu^{(ii)}(\omega)\;=\;
\frac{8 \pi^2}{3\omega}
\;,
\label{ee8}
\end{eqnarray}
where $y=\ln(1/x)$.
The saddlepoint $\omega_S^{(i)}$ of the integrand in (\ref{ee8}) is
determined by the condition
\begin{equation}
\left.\frac{d}{d\omega} \left\{\omega \,y \;+ \;\nu^{(i)} \,
\hat t\right\}\right|_{\omega =\omega^{(i)}_S}
\;=\;0
\end{equation}
and similarly  $\omega_S^{(ii)}$.
Using the method of steepest descent then gives the following results.

\noindent
{\bf (i)} In the region $Q^2 \ge k_\perp^2 \ge Q_0^2$:
\begin{eqnarray}
\left. x \,g(x,k_\perp^2)\right|_{Q^2 \ge k_\perp^2 \ge Q_0^2}
&=&
N_1(x)\;+\; G_1(x,k_\perp^2)\;\exp\left[-\frac{11 C_A}{12 \pi b}
H_1(k_\perp^2)\right]
\;\exp\left[\sqrt{\frac{4 C_A}{\pi b}
H_1(k_\perp^2)\ln\left(\frac{1}{x}\right)}\right]
\nonumber \\
& &\nonumber \\
N_1(x)&=& x g(x,k_\perp^2=Q^2)\;=\; \delta(1-x) \delta(k_\perp^2-Q^2)
\nonumber \\
G_1(x,k_\perp^2)&=& \frac{1}{\sqrt{4\pi}}\;\left[\frac{C_A}{\pi b}
\,H_1(k_\perp^2)\right]^{1/4}
\;\left[\ln\left(\frac{1}{x}\right)\right]^{-3/4}
\label{gsol1}
\\
H_1(k_\perp^2)&=&
\ln\left[\frac{\ln(Q^2/\Lambda^2)}{\ln(k_\perp^2/\Lambda^2)}\right]
\;.
\nonumber
\end{eqnarray}

\noindent
{\bf (ii)} In the region $Q_0^2 \ge k_\perp^2 \ge Q_0^2$:
\begin{eqnarray}
\left. x \,g(x,k_\perp^2)\right|_{Q_0^2 \ge k_\perp^2 \ge \Lambda^2}
&=&
N_2(x)\;-\; G_2(x,k_\perp^2)\;
\exp\left[\sqrt{\frac{4\rho}{3}
H_2(k_\perp^2)\ln\left(\frac{1}{x}\right)}\right]
\nonumber \\
& &\nonumber \\
N_2(x)&=& x g(x,k_\perp^2=Q_0^2)
\nonumber \\
G_2(x,k_\perp^2)&=& \frac{1}{\sqrt{4\pi}}\;\left[\frac{\rho}{3}
\,H_2(k_\perp^2)\right]^{1/4}
\;\left[\ln\left(\frac{1}{x}\right)\right]^{-3/4}
\label{gsol2}
\\
H_2(k_\perp^2)&=&
\frac{1\,-\,(Q_0^2/k_\perp^2) \left(\frac{}{}1 - \ln (Q_0^2/k_\perp^2)\right)}
{(Q_0^2/\Lambda^2)\ln (Q_0^2/\Lambda^2)}
\;.
\nonumber
\end{eqnarray}

In Fig. 7 the $x$-spectra of (\ref{gsol1}) and (\ref{gsol2}) are shown
for different values of $k_\perp^2$ with fixed $Q_0=1$ GeV. Two different
initial distributions were chosen to start the evolution from $k_\perp^2=
(3.5\,\mbox{GeV})^2$,
one flat in rapidity and the other one a Gaussian form.
The parameter $\rho$ introduced in (\ref{g2x}) was set equal to one.
For $k_\perp^2\,\lower3pt\hbox{$\buildrel > \over\sim$}\, Q_0^2$
the gluon distribution $xg(x,k_\perp^2)$ is just the
well known LLA solution with its strong increase at small $x$ as $k_\perp^2$
decreases.
For $k_\perp^2<Q_0^2$  however,  the effect is reversed such that
$g(x,k_\perp^2)$ decreases as $k_\perp^2$ falls below $Q_0^2$. This
suppression, which is particularly substantial at small $x$, reflects the
"condensation"
of gluons in the collective background field $\chi$.
\smallskip

The $k_\perp^2$-dependence of the total gluon multiplicity
can also be estimated in the above approximation.
The gluon multiplicity
is given by the $\omega =0$ moment,
\begin{equation}
N_g(k_\perp^2)\;=\;g(\omega =0,k_\perp^2)\;=\;\int_0^1 dx \;g(x,k_\perp^2)
\;.
\label{Ng}
\end{equation}
Using eqs. (\ref{ee5})-(\ref{A}),
one arrives in the soft limit ($z\ll 1$) \cite{furmanski79,basetto80}
at the following integral equations
that govern  the approximate behaviour
of the gluon multiplicity for the two cases (i) and (ii):
\begin{eqnarray}
(i)& &
k_\perp^2 \frac{\partial}{\partial k_\perp^2} \,N_g(k_\perp^2) \;=\;
\frac{\alpha_s(k_\perp^2)}{2\pi}\;\left\{
2\,C_A\int_{k_\perp^2}^{Q^2}\frac{d k^{'\,2}}{k^{'\,2}}\,
N_g(k_\perp^{'\,2})\;-\;\frac{1}{2}\; N_g(k_\perp^2)
\right\}
\nonumber
\\
& & \nonumber \\
(ii)& &
k_\perp^2 \frac{\partial}{\partial k_\perp^2} \,N_g(k_\perp^2) \;=\;
-\,\frac{\lambda_\chi(k_\perp^2)}{2\pi}\;
\left( 8 \pi^2 \, c_{gg\rightarrow \chi} \frac{\Lambda^2}{k_\perp^2}
\right)\;
\int_{k_\perp^2}^{Q_0^2}\frac{d k^{'\,2}}{k^{'\,2}}\,
N_g(k_\perp^{'\,2})
\label{ee9}
\;.
\\ & & \nonumber
\end{eqnarray}
On account of the $k_\perp^2$-dependence of $\alpha_s$ (\ref{alpha})
and $\lambda_\chi$ (\ref{lambda2}), the corresponding solutions are obtained
as:
\begin{eqnarray}
(i)& &
\left. N_g(k_\perp^2)\right|_{Q^2 \ge k_\perp^2 \ge Q_0^2}
\;=\;
N_g(Q^2)\;\left(\frac{\ln(Q^2/\Lambda^2)}{\ln(k_\perp^2/\Lambda^2)}\right)^{-1/4}
\;
\frac{\exp\left[2\,\sqrt{(C_A/\pi b)\,\ln(Q^2/\Lambda^2)}\right]}
{\exp\left[2\,\sqrt{(C_A/\pi b)\,\ln(k_\perp^2/\Lambda^2)}\right]}
\nonumber
\\
& & \nonumber \\
(ii)& &
\left. N_g(k_\perp^2)\right|_{Q_0^2 \ge k_\perp^2 \ge \Lambda^2}
\;=\;N_g(Q_0^2)\; \exp\left[-\,\frac{\rho}{2}\,\frac{\Lambda^2}{k_\perp^2}\,
\frac{1 + [1+\ln(k_\perp^2/Q_0^2)]^2 - 2
k_\perp^2/Q_0^2}{\ln(Q_0^2/\Lambda^2)}\right]
\;.
\label{ee10}
\end{eqnarray}
In the region (i) $Q^2 \ge k_\perp^2 > Q_0^2$, the multiplicity coincides
with the QCD result \cite{basetto83}, characterized by a rapid
growth as the gap between the
hard scale $Q^2$ and $k_\perp^2$ increases.
On the other hand, in  the region
(ii) $Q_0^2 \ge k_\perp^2 > \Lambda^2$, the multiplicity becomes strongly
damped.
The exponent is always negative so that the number of gluons
rapidly decreases and vanishes at $k_\perp^2=0$, ensuring that no
gluons and therefore no color
fluctuations exist at distances
$R \,\lower3pt\hbox{$\buildrel > \over\sim$}\, \Lambda^{-1}$.
This behaviour is evident in Fig. 8, where $N_g(k_\perp^2)$ is plotted
versus $\Lambda^2/k_\perp^2$, starting from $\Lambda^2/Q^2\ll 1$.
\smallskip

It must be emphasized that the {\it perturbative} evolution of gluons is
cut off at $Q_0^2$, and in  the transition region
$Q_0^2 \ge k_\perp^2 > \Lambda^2$ the evolution is purely {\it
non-perturbative}, although it is
described here as an extension of the perturbative evolution above $Q_0^2$
and treated on the same footing.
\bigskip

\noindent {\bf 3.3 Flux tube configurations of
gluons interacting with the mean field {\boldmath $\chi$}}
\medskip

In Secs. 3.1 and 3.2 the evolution of the gluon configuration between the
fragmenting $q\bar q$ pair was analyzed in terms of the non-perturbative
modification as a function of $\lambda_\chi(k_\perp^2) =
\tilde{\xi}_\chi^2/(4\pi)$.
However, the coupling strength $\lambda_\chi$ or $\tilde{\xi}_\chi$ between
the gluon field and the $\chi$-field must be determined by the dynamics itself,
since $\tilde{\xi}_\chi$ is the Fourier transform of the $\chi$-dependent
coupling function $\xi(\chi)$ in eq. (\ref{LQCDchi}).

The dynamics is governed by the the  coupled system of equations
eqs. (\ref{e1})-(\ref{e3}),
which  can now be solved numerically
by utilizing the definition (\ref{gdef}) together with the general solution for
the
gluon distribution given by (\ref{sol1}) and (\ref{sol2}).
To do so, one obtains the expectation value of eq. (\ref{e2})
by multiplying with the multigluon state vector $\langle P|$ and $|P\rangle$
from the left and right, respectively.
In the mean field approximation, $\chi = \bar \chi$ is a $c$-number function,
so that this operation affects only the $F_{\mu\nu}F^{\mu\nu}$
term which gives the gluon distribution by virtue of (\ref{gdef}).
With the explicit solution of the gluon distribution (\ref{gsol1}) and
(\ref{gsol2})
inserted, the remaining task
is to solve  the single equation (\ref{e2}) for the $\chi$-field.
(Recall that the equation (\ref{e3}) for the $U$-field couples only to
$\chi$ and is in principle readily solved once the solution for $\chi$ is
known.)

An interesting phenomenological application
\cite{wilets,bickeboeller2,grabiak91}
is to calculate the
{\it string tension} $t$, which characterizes the linearly rising potential
between the  $q\bar q$-pair due to the gluon interactions.
usually obtained by fitting heavy quarkonium spectra \cite{qqspectra}
with a non-relativistic potential of the form
\begin{equation}
V_{q\bar q}(r)\;=\;-\frac{a}{r}\;+\; t\;r^2
\label{Vqq}
\;,
\end{equation}
where $a = 4 \alpha_s/3$. Typical fit values for the string constant $t$
range from 750 MeV/fm to 950 MeV/fm.

Here I shall estimate the string constant within the
present approach on the basis of the
equations of motion (\ref{e1})-(\ref{e3}). Similar calculations
have been done earlier in the framework of the static MIT bag model
\cite{MIT} and the Friedberg-Lee soliton model \cite{bickeboeller2,grabiak91}.

In the following I will consider the fragmentation of a heavy $q\bar q$ pair
within an {\it adiabatic} approximation. That is, a quasistatic treatment is
employed which neglects the motion of the $q\bar q$ pair and considers the
instantanous
gluon configuration in between the pair.  This should provide is a reasonable
approximation,
because one  can view the $i^{th}$ gluon as being emitted from
the $q\bar q$ pair plus gluons $g_1, g_2, \ldots , g_{i-1}$, with the
spatial coordinates of these "sources" being frozen during the emission
of the gluon $i$ \cite{mueller94}.  In a  space-time picture of the
fragmentation of
the $q\bar q$ and its emitted gluons it is the change of the typical transverse
momenta $k_\perp$ or
transverse separation $r_\perp\propto k_\perp^{-1}$ of gluons
which governs the dynamical transition from short distance, unconfined regime
to
long distance, confined stage \cite{pQCD2}, because there must be a critical
separation of color charges
beyond which the total color is screened. In the present case, the role
of this non-perturbative phenomenon is played by the $\chi$-field.
In the adiabatic approximation, there is no explicit dependence
on the longitudinal variables here, because the separation in the
transverse plane is independent of when during the cascade,
or where along the jet axis, a gluon was produced \cite{kogut73}.

Thus, within the adiabatic treatment, one may use the separation $R_{q\bar q}$
of the receding $q\bar q$ as a measure for
the typical gluon transverse momenta $k_\perp^2\approx R_{q\bar q}^{-2}$.
By minimizing the energy per unit length of this system,
one obtains then for each gluon configuration at a given $R_{q\bar q}$
the form of $\chi$ and the string tension $t$.
The total energy per unit length, the string tension $t$,
receives contributions from both the $\chi$-field in the low energy regime and
the
gluon field in the high energy domain, which on account of the
the trace anomaly (\ref{theta}) is given by:
\begin{equation}
t \;\equiv\; \frac{E}{R_{q\bar q}}\;=\; \int d A  \left[
\,\frac{1}{2}\left|\nabla \chi \right|^2
\;+\; V(\chi)\,\right] \;\,+\;\,
\int dA \; \kappa(\chi)\;\langle P \vert \frac{\beta(\alpha_s)}{4\alpha_s}\,
\,F_{\mu\nu, a} F^{\mu\nu}_a \vert P\rangle
\;.
\label{t}
\end{equation}
Here $A$ is the cross-sectional area of the flux tube of the
gluons between the $q\bar q$, perpendicular to $R_{q\bar q}$.
In one loop order the beta-function is $\beta= -b \alpha_s^2$ with $b=
33/(12\pi)$ for
$N_f=0$.
Then, by using eq. (\ref{gdef}) to express the second integral in terms of the
gluon
distribution $g(x,k_\perp^2)$, assuming cylindrical symmetry along
the $R_{q\bar q}$ axis, and minimizing $t$ with respect to $\chi$,
one arrives at the following non-linear integro-differential equation
($r$ is the radial coordinate perpendicular to $R_{q\bar q}$):
\begin{equation}
-\,\left[ \frac{d^2}{dr^2} + \frac{1}{r} \frac{d}{dr}\right]
\;\chi(r) \;+\; \frac{\partial V(\chi)}{\partial \chi} \;
+ \;(P^+)^2\;\frac{b}{2}\;I_g(R_{q\bar q})\;
\int dr \,r\, \kappa(\chi)
\;=\;0\;,
\label{t1}
\end{equation}
where
\begin{equation}
I_g(R_{q\bar q})\;=\;
\int_{1/R_{q\bar q}}^{(P^+)^2} d^2 k_\perp \,\alpha_s(k_\perp^2)\,
\int_0^1 dx\,x\,g(x,k_\perp^2)
\;.
\label{I}
\end{equation}
In pulling $I_g$ out of the $r$-integral, it is assumed that the spatial
distribution
of gluons is approximately homogenous.
Eq. (\ref{t1})  determines the energetically most favorable flux tube
configuration. However, a physical meaningful
flux tube solution has to satisfy the {\it constraint}
that the system as a whole, $q\bar q$ plus gluons,
must form a global color singlet, implying that all of the color flux that
originates from the $q$ must be directed towards the $\bar q$.
In other words, the total color electric flux
through a plane between the $q$ and $\bar q$
must be equal the color charge $Q_q= g_s T_a$ on one of them
\cite{bickeboeller2,grabiak91}.
This translates into the requirement that the  gluons in the flux tube streched
between
$q$ and $\bar q$ with certain $R_{q\bar q}$ carries a total color charge
squared that is equal to
$Q_q^2$, the one of $q\bar q$. Define
\begin{equation}
\phi_g \;=\; \frac{(P^+)^2}{A} \int d^2r \;J_g(R_{q\bar q})
\;,
\label{flux}
\end{equation}
where $A$ is the cross-sectional area of the flux tube, and
\begin{equation}
J_g(R_{q\bar q})\;=\;
\int_{1/R_{q\bar q}}^{(P^+)^2} d^2 k_\perp \int_0^1 dx\,g(x,k_\perp^2)
\;,
\label{J}
\end{equation}
is the total number of gluons radiated from the initial point of $q\bar
q$-production up to $R_{q\bar q}$.
Then the above constraint then reads
\begin{equation}
\phi_g \;\stackrel{!}{=} \;
Q_q^2\;=\; \langle\, g_s^2 \; T_a \cdot T_a \,\rangle
\;=\;\left. \frac{16\pi}{3}\;\alpha_s(k^2)\right|_{k^2=R_{q\bar q}^{-2}}
\label{constraint}
\;.
\end{equation}
Combining eqs. (\ref{t1})-({\ref{constraint}), one arrives at
\begin{equation}
-\,\left[ \frac{d^2}{dr^2} + \frac{1}{r} \frac{d}{dr}\right]
\;\chi(r) \;+\; \frac{\partial V(\chi)}{\partial \chi} \;
+ \;\frac{8\pi}{3} \,b\,\alpha_s(R_{q\bar q}^{-2}) \;\frac{I_g(R_{q\bar
q})}{J_g(R_{q\bar q})}
\;\int dr \,r\, \kappa(\chi)
\;=\;0\;,
\label{t2}
\end{equation}
which is now independent of the overall boost momentum $(P^+)$, and thus of the
initial hard scale $Q^2=(P^+)^2$, as it
should be.

Solving eq. (\ref{t2}) numerically \cite{colsys},
subject to the the boundary conditions $\chi ^\prime(0)=0$ and
$\chi(\infty)=\chi_0$
yields the solutions for $t$, $\chi$ and $\kappa(\chi)$ shown in Fig. 9.
The reasonable parameter values \cite{ellis90} $\chi_0=f_\pi$ and bag constant
$B=(150\,\hbox{MeV})^4$
were chosen for $V(\chi)$, eq. (\ref{V}), and  the coupling function $\kappa$
was taken of the form (\ref{kappa2}).
One sees that with increasing separation $R_{q\bar q}$ of $q$ and $\bar q$, the
gluons first multiply which
results in a growing string tension,
but then gluon condensation sets in, yielding a saturating behaviour of $t$
with the string constant approaching $t\simeq 1$ GeV/fm (Fig. 9a).
For $R_{q\bar q}\approx 1$ fm the gluon field is completely confined
within a flux tube of radius $r\simeq 1$ fm (Fig. 9b).
For comparison, a simple estimate within the MIT bag model \cite{MIT}
gives $t = 910$ MeV/fm, but a rather large tube radius of 1.6 fm.
Detailed calculations with in the soliton model \cite{bickeboeller2} gave
qualitatively similar results.
Finally, Fig. 10 shows in correspondance to Fig 9b the
form of the potential $V(\chi)\equiv V(\chi,U)|_{m_q=0}$, defined in eq.
(\ref{V}),
and the effective squared mass $M^2(\chi) = d^2 V(\chi)/d \chi^2$ for the
same parameter values as above.
\bigskip

\noindent {\bf 4. SUMMARY AND OUTLOOK}
\bigskip

The effective QCD field theory approach presented here to describe
the dynamics of high energy partons in the presence of a collective confinement
field
provides a framework that has the potential to be developed towards a
systematic
description of the hadronization mechanism.
The corresponding effective action has been constructed such that it
\begin{itemize}
\item[(i)]
incorporates both parton and hadron degrees of freedom;
\item[(ii)]
recovers the exact QCD (Yang-Mills) action with its symmetry properties at
short
space-time distances;
\item[(iii)]
merges into an effective low energy description of hadronic degrees of freedom
at large
distances;
\item[(iv)]
allows for a dynamical description of parton-hadron conversion on the basis
of the resulting equations of motion.
\end{itemize}
As an exemplary demonstration, the approach was applied to the evolution of a
fragmenting $q\bar q$ pair with its generated gluon distribution, starting from
a large hard scale $Q^2$ all the way downwards to $\Lambda^2$.
The transformation of the initially high virtual gluons to a gluon condensate
field $\chi$
was studied in terms of the coupled evolution of the gluon distribution and the
mean field $\chi$. The solution of the equations of motion yields color flux
tube
configurations with an associated energy per unit area (string tension) of
about
1 GeV/fm, consistent with the common estimates.
\smallskip

In perspective, important points to be addressed in the future, are:
\begin{itemize}
\item [(i)]
The establishment of the relation with the exact renormalization group
equation for the effective action as derived by Reuter and Wetterich
\cite{reuter93}
is desirable. This would allow to quantify  the effect of consecutively
integrating
out all quantum fluctuations of gluons and quarks with momenta larger than some
infra-red
cut-off scale $Q_0$, the variation of which determines the confinement
dynamics.
\item [(ii)]
With the inclusion of quark degrees of freedom and possibly quantum
fluctuations of the $\chi$-
and $U$-fields, one could calculate e.g. the mass spectrum of glueball and
meson
excitations as physical hadrons.
This would provide a complete description from a physical initial state, via
a not directly observable deconfined partonic stage, up to the formation of
observable hadronic excitations.
\item [(iii)]
Ultimately one would like to
address the microscopic dynamics in full 6-dimensional phase-space \cite{pcm},
with explicit inclusion of the color degrees of freedom and the local color
stucture.
This could be realized in a transport theoretical formulation similar as in
Ref. \cite{kalmbach93}, in which the partons
propagate with a modified propagator that embodies the effects of the mean
field
$\chi$ in the effective mass. As the confining field becomes significant
the effective mass increases and asymptotically becomes infinite so that
the propagation of color fluctuations ceases.
\item [(iv)]
The possible applications are manifold. One particular interest is the
expected (non)equilibrium QCD phase transition in high energy systems
as in heavy ion collisions or the early universe, an issue which
could be addressed along the lines of Campbell, Ellis and Olive  \cite{ellis90}
in combination with the space-time evolution of the multi-parton system
\cite{pcm}
in the presence of the collective field.
\end{itemize}
\bigskip

\section*{Acknowledgements}
\vspace{-2mm}
\noindent
I thank  J. Ellis, H.-T. Elze and G. Veneziano for valuable suggestions. Thanks
also
to\\ L. Wilets and W. Hoefer for providing the computer code for the flux tube
calculation.

\newpage
\noindent {\bf APPENDIX A}
\bigskip

For completeness a "derivation" of the evolution equation
(\ref{ee0}) is given in the spirit of Lipatov \cite{lipatov74}.
The full propagator of a single gluon of momentum $k$ may be represented as
\begin{equation}
D_{\mu \nu}(k)\;=\;  \;\frac{d_{\mu\nu}(k)}{k^2 \,
\left(\frac{}{}1 + \Pi(k^2) \right)} \;\equiv \;
\frac{D^0_{\mu \nu}(k)}{1 + \Pi(k^2)}
\;,
\label{prop0}
\end{equation}
where $D^0_{\mu \nu}= d_{\mu\nu}/k^2$ is the free propagator, and
with the choice of gauge for the gluon fields $\eta \cdot A = 0$,
\begin{equation}
d_{\mu \nu}(k) \;=\; g_{\mu \nu} \;-\;
\frac{k_\mu \eta_\nu + k_\nu\eta_\mu}{k\cdot \eta}
\;\;\;\;,\;\;\;\;\;\;\;\;\;\;\;
d_{\mu\mu}(k) \;=\; 2
\label{prop1}
\end{equation}
and $k^\mu d_{\mu\nu}(k)|_{k^2=0}=0$
guarantees that only 2 physical (transverse) gluon polarizations
propagate on mass shell.
The  self-energy part
\begin{equation}
\Pi(k^2) \;=\; \Pi_{g\leftrightarrow g}(k^2) \;+\;\Pi_{g\leftrightarrow
\chi}(k^2)
\label{Pi}
\end{equation}
contains both the gluon self-interactions and the "medium" corrections
due to the coupling to the confining background field $\chi$
Expanding
$\Pi_{g\leftrightarrow g}$ and $\Pi_{g\leftrightarrow \chi}$ in powers
of the squared couplings $g_s^2$ and $\xi_\chi^2$, respectively,
the contribution to one loop order is determined by the
total gluon "decay" probability, i.e. the probability of losing
a gluon out of a momentum space element between $k^2$ and $k^2+dk^2$:
\begin{equation}
w_g(k^2)\;=\;
\frac{\partial}{\partial \ln(k^{\prime \,2}/\Lambda^2)} \,
\left[
\Pi_{g\leftrightarrow g}(k^{\prime\,2}) +\Pi_{g\leftrightarrow \chi}(k^{\prime
\,2})
\right]_{k^{\prime\,2} = k^2}
\;\,\equiv \;\,
w_{g\leftrightarrow g}\;+\; w_{g\leftrightarrow \chi}
\;.
\label{wg}
\end{equation}
Here, $w_{g\leftrightarrow g}$ and $w_{g\leftrightarrow \chi}$ are the
inclusive
probabilities for a gluon to emit (absorb) another gluon, due to the
self-interaction,
respectively the interaction with the $\chi$-field, corresponding to the
diagrams in Fig. 4,
\begin{eqnarray}
w_{g\leftrightarrow g}(k^2) &=&
\;\int_0^1 dx
\int_x^1 \frac{dx'}{x'} \;\left[ w_{g\rightarrow gg}(x',x,k^2) \,+\,
w_{gg\rightarrow g}(x',x,k^2) \right]
\nonumber
\\
w_{g\leftrightarrow \chi}(k^2) &=&
\;\int_0^1 dx
\int_x^1 \frac{dx'}{x'} \;\left[ w_{g\rightarrow g\chi}(x',x,k^2) \,+\,
w_{gg\rightarrow \chi}(x',x,k^2) \right]
\;.
\label{wgx}
\end{eqnarray}
The individual contributions in square brackets $w(x_1,x_2,k^2)$ can be
obtained
in the standard fashion \cite{altarelli77,amati78} by evaluating the
cross-section ratios (c.f. Appendix B)
\begin{equation}
\frac{k_\perp^2}{\sigma^{(0)}}\,\frac{d\sigma^{(1)}}{d z dk_\perp^2}
\;=\;
\frac{g^2}{8\pi^2}\;
\gamma_{a\rightarrow bc}(z)
\label{gamma2}
\;,
\end{equation}
where $g$ denotes the appropriate coupling for the process under consideration
(here $g=g_s$ or $g = \xi_\chi$),
$\sigma^{(0)}$ is the Born cross-section for the production of
a gluon $a$ and $\sigma^{(1)}$ represents the first order correction
associated with the "decay" $a\rightarrow b c$.

For the process $g\rightarrow gg$ and its reversal $gg \rightarrow g$
the probability distributions (\ref{gamma}) are well known
\cite{altarelli77,close89}.
Assigning the momentum fractions as
$x_1\rightarrow x_2, (x_1-x_2)$ for $g\rightarrow gg$ and
$x_1, (x_2-x_1)\rightarrow x_2$ for $gg\rightarrow g$, one has
\begin{eqnarray}
w_{g\rightarrow gg}(x_1,x_2,k^2) &=&
\frac{\alpha_s(k^2)}{2\pi}\;\gamma_{g\rightarrow
gg}\left(\frac{x_2}{x_1}\right)
\nonumber \\
w_{gg\rightarrow g}(x_1,x_2,k^2) &=&
\frac{\alpha_s(k^2)}{2\pi}\;
\left[ 8\pi^2 \,c_{gg\rightarrow g}\,\frac{\Lambda^2}{k^2}\;
\; \frac{x_1(x_2-x_1)}{x_2^2}\right]
\;\gamma_{g\rightarrow gg}\left(\frac{x_1}{x_2}\right)
\label{ggg}
\;,
\end{eqnarray}
where in the second expression the factors in square brackets
arise from the difference of phase-space and flux factors for fusions
compared to branchings.  The color factor is $c_{gg \rightarrow g}= 1/8$ and
\begin{equation}
\gamma_{g \rightarrow g g} (z) \;=\;
2\,C_A\;\left( z ( 1 - z ) + \frac{z}{1-z} + \frac{1-z}{z} \right)
\;,
\label{gammagg}
\end{equation}
where $C_A = N_c = 3$, and $z$ is the fraction of $x$-values of daughter to
mother gluons.

The new, additional processes  are the friction process
$g \rightarrow g \chi$, corresponding
to energy-momentum transfer from gluons to the $\chi$-field, and
the fusion process $gg \rightarrow\chi$, by which two gluons
couple color neutral to the $\chi$-field and "annihilate"
\footnote{Since the conversion of partons into hadrons in the process
of fragmentation is an irreversible process, the spontanous production of
gluons
by the $\chi$-field $\chi \rightarrow gg$, as well as the energy transfer
from the $\chi$-field to  the gluons, $g \chi \rightarrow g$, are
omitted. These latter interactions would counteract the transition,
which certainly is possible in the sense of local fluctuations, but
globally, and in the average, the parton-hadron conversion is a one-way process
in the present context.}.
As outlined in Appendix B, one arrives at
(with the assignment
$x_1\rightarrow x_2, (x_1-x_2)$ for $g\rightarrow g\chi$ and
$x_1, (x_2-x_1)\rightarrow x_2$ for $gg\rightarrow \chi$)
\begin{eqnarray}
w_{g\rightarrow g \chi}(x_1,x_2,k^2) &=&\;
\frac{\lambda_\chi(k^2)}{2\pi}\;
\gamma_{g\rightarrow g\chi}\left(\frac{x_2}{x_1}\right)
\nonumber \\
w_{gg \rightarrow \chi}(x_1,x_2,k^2) &=&
\frac{\lambda_\chi(k^2)}{2\pi}\;
\left[ 8\pi^2 \,c_{gg\rightarrow g}\,\frac{\Lambda^2}{k^2}\;
\; \frac{x_1(x_2-x_1)}{x_2^2}\right]
\;\gamma_{\chi\rightarrow gg}\left(\frac{x_1}{x_2}\right)
\label{ggc}
\;,
\end{eqnarray}
where $c_{gg\rightarrow \chi} = 1/8$, and
\begin{eqnarray}
\gamma_{g \rightarrow g \chi} (z) &=&
\frac{1}{4} \left(\frac{ 1\;+\;z^2}{1\;-\;z}\right)
\nonumber
\\
\gamma_{\chi \rightarrow g g} (z) &=&
8 \left(\frac{}{} z^2\, - \, z \, +\, \frac{1}{2}\right)
\label{gammagc}
\;.
\end{eqnarray}

The total interaction probability $w_g$ (\ref{wg})
determines via the unitarity condition (\ref{eenorm}) the Sudakov formfactor
$F_g$,
\begin{equation}
F_g(Q^2,k^2) \;=\;\exp \left[-\, \int_{k^2}^{Q^2}\frac{dk^{'\,2}}{k^{'\,2}}
w_g(k^{'\,2}) \right]
\,
\label{Fg2}
\end{equation}
which is the probability that a gluon does not at all interact
(i.e.  emit or absorb other gluons)
while degrading its virtuality from $Q^2$ to $k^2$.

The self-energy part (\ref{Pi}) is now readily evaluated on the basis of
eq. (\ref{wg}), and inserted in the representation
(\ref{prop0}), one obtains the single gluon propagator at one loop order,
\begin{equation}
D_{\mu\nu}(k)\;=\;D_{\mu\nu}^0(k) \;
\left[1 \;+\;
 \;\int_{k^2}^{Q^2}\frac{d k^{'\,2}}{k^{'\,2}}\,
\int_x^1 \frac{dx'}{x'} w(x',x,k^2)\right]
\;.
\end{equation}
The corresponding "jet calculus" \cite{konishi79}
Greens function $D_g(x,k^2;x_0,Q^2)$ of eq.
(\ref{ee0})
describes how a system of gluons evolves in the variable $x$ and the virtuality
$k^2$ through the gluon self-interactions and in the presence of
the confining background field $\chi$.  It is given by the convolution of
the single gluon propagator (\ref{prop0}) with the gluon distribution function
$g(x,k^2)$.
Defining
\begin{equation}
D_g(x,k^2;x_0,Q^2)\;\equiv\;  g_{\mu\nu}\, D^{\mu\nu}(k)\;\otimes\;g(x,k^2)
\;,
\end{equation}
the self-consistent iteration of one loop contributions to all orders
within the LLA gives the evolution equation for the gluon distribution
with respect to the variables $x$ and $k^2$:
\begin{eqnarray}
k^2 \frac{\partial}{\partial k^2} \,g(x,k^2)
&\equiv&
k^2 \frac{\partial}{\partial k^2} \,D_g(x,k^2;x_0,Q^2)
\nonumber \\
&=&
+\;\frac{\alpha_s(k^2)}{2 \pi} \;\left\{
\int_x^1 \frac{dx'}{x'} \, g(x',k^2) \;
\gamma_{g\rightarrow gg}\left(\frac{x}{x'}\right)
\;-\;\frac{1}{2} \,g(x,k^2)\, \int_0^1 dz \,
\gamma_{g\rightarrow gg}\left(z\right)
\right\}_{k^2 \ge Q_0^2}
\nonumber \\
& & -\;
\frac{\alpha_s(k^2)}{2 \pi} \;
\left(8\pi^2 \frac{\Lambda^2}{k^2}\right)\;
\left\{
\int_0^1 dx' \, g^{(2)}(x,x',k^2) \;\
\Gamma_{gg\rightarrow g}\left(x,x',x+x'\right)
\right.
\nonumber \\
& & \;\;\;\;\;\;\;\;\;\;\;\;\;\;
\;-\;
\left.
\frac{1}{2}\,
\int_x^1 dx' \, g^{(2)}(x-x',x',k^2) \;
\Gamma_{gg\rightarrow g}\left(x-x',x',x)\right)
\right\}_{k^2 \ge Q_0^2}
\nonumber \\
& & +\;
\frac{\lambda_\chi(k^2)}{2 \pi} \;\left\{
\int_x^1 \frac{dx'}{x'} \, g(x',k^2) \;
\gamma_{g\rightarrow g\chi}\left(\frac{x}{x'}\right)
\;-\;g(x,k^2)\, \int_0^1 dz \, \gamma_{g\rightarrow g\chi}\left(z\right)
\right\}
\nonumber \\
& & -\;
\frac{\lambda_\chi(k^2)}{2 \pi} \;
\left(8\pi^2\frac{\Lambda^2}{k^2}\right)\;
\left\{
\int_0^1 dx' \, g^{(2)}(x,x',k^2) \;
\Gamma_{gg\rightarrow \chi}\left(x,x',x+x'\right)
\right\}
\label{ee3}
\end{eqnarray}
where the factor 1/2 in the first (third) term arises from the
indistinguishability
of the two gluons emerging from (coming in) the branching (fusion) vertex.
The function
$g^{(2)}(x_1,x_2,k^2)$ denotes the 2-gluon density,
and the gluon fusion functions
$\Gamma$ are defined in accord with (\ref{ggg}) and (\ref{ggc})
as
\begin{equation}
\Gamma_{12\rightarrow 3}(x_1,x_2,x_3) \;=\; c_{12\rightarrow 3} \;\frac{x_1
x_2}{x_3^2} \;
\gamma_{3\rightarrow 12}\left(\frac{x_1}{x_3}\right)
 \;=\; c_{12\rightarrow 3} \;\frac{x_1 x_2}{x_3^2} \;
\gamma_{3\rightarrow 21}\left(\frac{x_2}{x_3}\right)
\;.
\label{Gamma}
\end{equation}
with $x_3=x_1+x_2$.
Changing to variables $(x,x_1,k^2) \rightarrow (x,z,k_\perp^2)$ and
using (\ref{Gamma}), one immediately arrives at eq. (\ref{ee1}).
\bigskip

\noindent {\bf APPENDIX B}
\bigskip

Here I will outline
the explicit calculation of the interaction probability densities
$\gamma_{g\rightarrow g \chi}$ and $\Gamma_{gg \rightarrow \chi}$
that appear in the evolution equation (\ref{ee3}) or (\ref{ee2})
in addition to the usual probability densities $\gamma_{g\rightarrow gg}$
and $\Gamma_{gg \rightarrow g}$.
Notice that
the vertex corresponding to three gluons coupling to the $\chi$-field
is unphysical and therefore to be excluded, because $\chi$ is
required to be a color singlet field.
On the other hand, the coupling of four gluons to $\chi$ is possible, however
in the LLA such diagrams are kinematically suppressed and can be neglected
\cite{pQCD2}.

Let $\sigma^{(N)}$ denote the spin and color averaged cross-section for the
production of a gluon at order $N$ in perturbation theory.
The probability distribution $\gamma_{a\rightarrow bc}$ in the
variable $z=x_b/x_a$
for the emission of a gluon $b$ in the process $a\rightarrow b c$,
is the given by the ratio of cross-sections
\begin{equation}
\frac{1}{\sigma^{(0)}}\,\frac{d\sigma^{(1)}}{d z}
\;=\;\frac{g^2}{8\pi^2}\;\gamma_{a\rightarrow bc}(z) \;\frac{d
k_\perp^2}{k_\perp^2}
\label{s1s0}
\;,
\end{equation}
where g is the appropriate coupling of the process,
$\sigma^{(0)}$ is the lowest order cross-section for the production of
a gluon $a$ and $\sigma^{(1)}$ represents the first order correction associated
with
the "decay" $a\rightarrow b c$.
The vertex function associated with the general $gg\chi$ coupling
is easily obtained from the interaction Lagrangian $\cal{L}[\psi,A,\chi]$, eq.
(\ref{LQCDchi}),
as:
\begin{equation}
V^{ab}_{\mu\nu}(k_1,k_2,k)\;=\;
-\tilde{\xi}_\chi(k)\;\delta^{ab} \;\left[\frac{}{}
(k_1\cdot k_2) g_{\mu\nu} \,-\, (1-a) \,k_{1\,\mu}k_{2\,\nu}\right]
\end{equation}
where $\tilde{\xi}_\chi(k^2)$ denotes the Fourier transform of
$\xi(\chi)$ in coordinate space,
$k_1$, $k_2$ are the gluon momenta and the convention is that all  four-momenta
are directed into the vertex.
The process $g\rightarrow g \chi$ gives then (setting $a=1$):
\begin{equation}
\sigma^{(1)}(k_1^2) \;=\;
\int \frac{d^3 k}{(2\pi)^3 2 k_0} \;\frac{2}{(k_1\cdot k)^2}
\sigma^{(0)}(k_1^2)\;
|\overline{\cal M}|^2
\;,
\end{equation}
where
\begin{equation}
|\overline{\cal M}|^2\;=\;
\frac{1}{16} \sum_{a,a',b,b'} \sum_{s_1,s_2} \,
V_{\mu\nu}^{ab} V^{\ast\,a' b'}_{\mu '\nu '}\,
e^{\ast\,\mu}(s_1) e^{\ast\,\nu}(s_2) e^{\mu '}(s_1) e_2^{\nu '}(s_2)
\;,
\end{equation}
where the factor 1/16 in front results from the averaging over initial
2 transverse polarizations and 8 color degrees, and it is summed over final
color and
spin polarizations $s_i$.
The sum over gluon polarizations $s_1, s_2$ must be performed over transverse
polarizations
only. This is achieved by the projection
\begin{equation}
\sum_{s_i} e^\mu(s_i)e ^{\ast \,\nu}(s_i)\;=\;
- g^{\mu\nu}\;+\;\frac{k^\mu k_i^\nu + k^\nu k_i^\mu}{(k\cdot k_i)}
\;-\;\frac{k^2\,k_i^\mu k_i^\nu}{(k\cdot k_i)^2}
\;.
\end{equation}
Assigning the momenta $k^\mu=(k^+,k^-,\vec k_\perp)$
of incoming (outgoing) gluon $k_1$ $(k_2)$
and the momentum $k$ transferred to the $\chi$-field as
\begin{eqnarray}
k_1&=& \left(\frac{}{}k_1^+,0,\vec 0_\perp\right)
\nonumber\\
k_2&=& \left((1-z) k_1^+,\frac{k_\perp^2}{(1-z)k_1^+},-\vec k_\perp\right)
\nonumber\\
k&=& \left(z k_1^+,\frac{k_\perp^2}{zk_1^+},\vec k_\perp\right)
\;,
\end{eqnarray}
and carrying out the appropriate change of integration variables,
the result is
\begin{equation}
\sigma^{(1)}(k_1^2) \;=\;
\sigma^{(0)}(k_1^2) \;
\frac{ \tilde{\xi}_\chi^2}{8 \pi^2}
\int_{k_0^2}^{k_1^2}\frac{d k_\perp^2}{k_\perp^2}\;
\int dz \;\left[ \frac{1}{4} \;\left( 1 - z \;+\; 2
\,\frac{z}{1-z}\right)\right]
\;.
\end{equation}
Hence, one can read off
\begin{equation}
\gamma_{g\rightarrow g\chi}(z) \;=\; \frac{k_\perp^2}{\sigma^{(0)}}\,
\frac{d\sigma^{(1)}}{d z dk_\perp^2}
\;=\;\frac{1}{4} \left(\frac{ 1\,+ \,z^2}{1-z}\right)
\;.
\end{equation}

In complete analogous manner the process $gg\rightarrow \chi$ can be
calculated. The procedure is to evaluate $\chi\rightarrow gg$, with
incoming momentum $k$ and the outgoing momenta $k_1$ and $k_2$.
Using the formula (\ref{Gamma}) one obtains the 2-gluon fusion
function for the reverse process $gg\rightarrow \chi$. The result is:
\begin{eqnarray}
\Gamma_{gg\rightarrow \chi}(x_1,x_2,x_3) &=&
c_{gg\rightarrow \chi} \;\frac{x_1 x_2}{x_3^2} \;
\gamma_{\chi\rightarrow gg}\left(\frac{x_1}{x_3}\right)
\;\;,\;\;\;\;\;\;\;\;\;
(x_3=x_1+x_2)\;,
\\
& &\nonumber \\
\gamma_{\chi\rightarrow gg}(z)
&=& 8\left( z^2\;-\;z \;+\;\frac{1}{2}\right)
\;.
\end{eqnarray}


\newpage

{\bf FIGURE CAPTIONS}
\bigskip

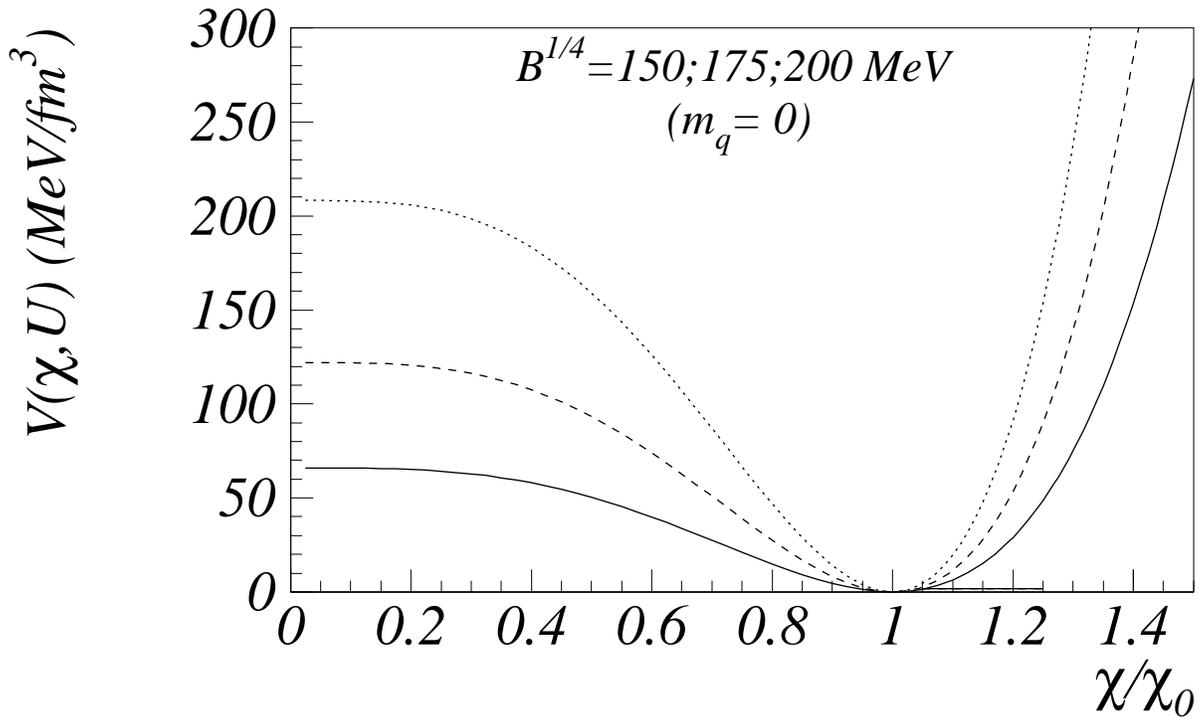
\noindent {\bf Figure 1:}
Typical shape of $V(\chi,U)$, eq. (\ref{V}), with
$\partial V/\partial \chi = 0$ at $\chi=\chi_0$
and $V=B$ at $\chi=0$, where $B=b\chi_0^4/4$ is the bag constant.
\bigskip

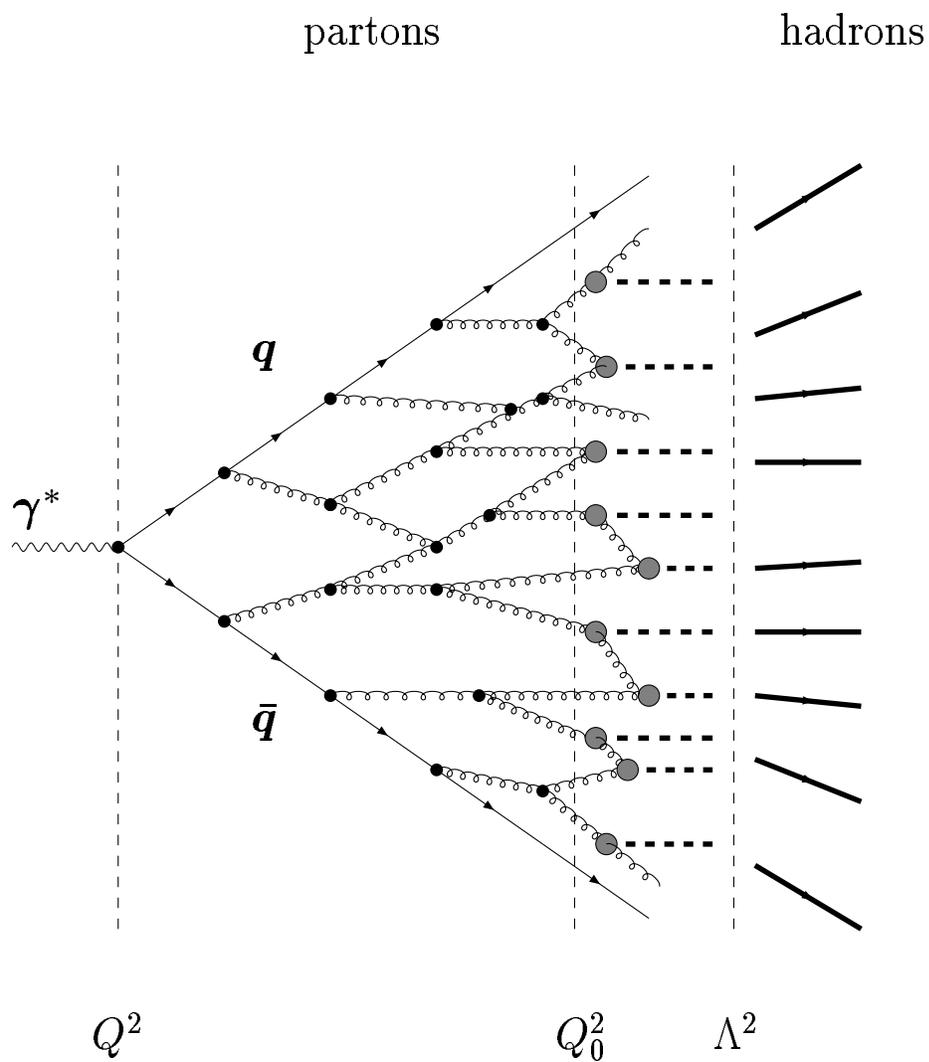
\noindent {\bf Figure 2:}
Schematics of the parton shower evolution of a fragmenting $q\bar q$ pair
with its gluon configuration as the virtualities
of the partons gradually degrade, starting from the hard scale $Q^2$.
At large gluon virtualities $k^2$ the shower develops
by perturbative branching processes,
but at $k^2 \simeq Q_0^2$ non-perturbative fusion and friction
processes set in, such that at $k^2 =\Lambda^2$ no colored
fluctuations remain.
\bigskip

\noindent {\bf Figure 3:}
Diagrammatic representation of
the  two-point Greens function of gluons, including both the
gluon (self) interactions and
the effective interaction with the confining background field $\chi$
(indicated by the dashed lines).
This gluon propagator
describes the evolution of a gluon from a chosen cascade branch in
$x$ and $k^2$, starting from $x_0$ and $Q^2$.
\bigskip

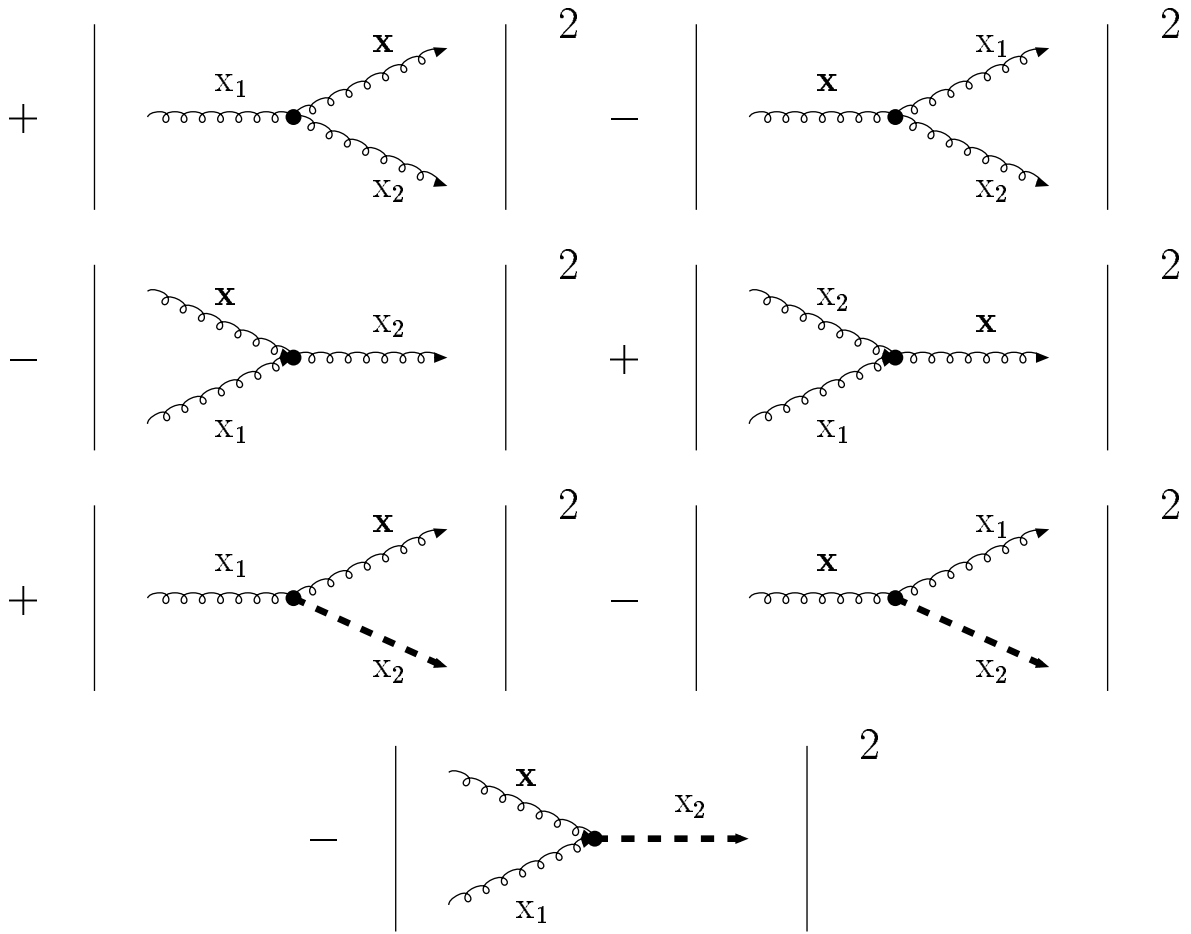
\noindent {\bf Figure 4:}
Diagrams for the interaction probabilities that determine the evolution
of the gluon distribution according to (\ref{ee2}). In the probabilistic
interpretation terms with positive signs leadt to a gain of gluons and terms
with negative sign to a loss.
The gluon with momentum fraction $x$ is the "observed" particle.
\medskip

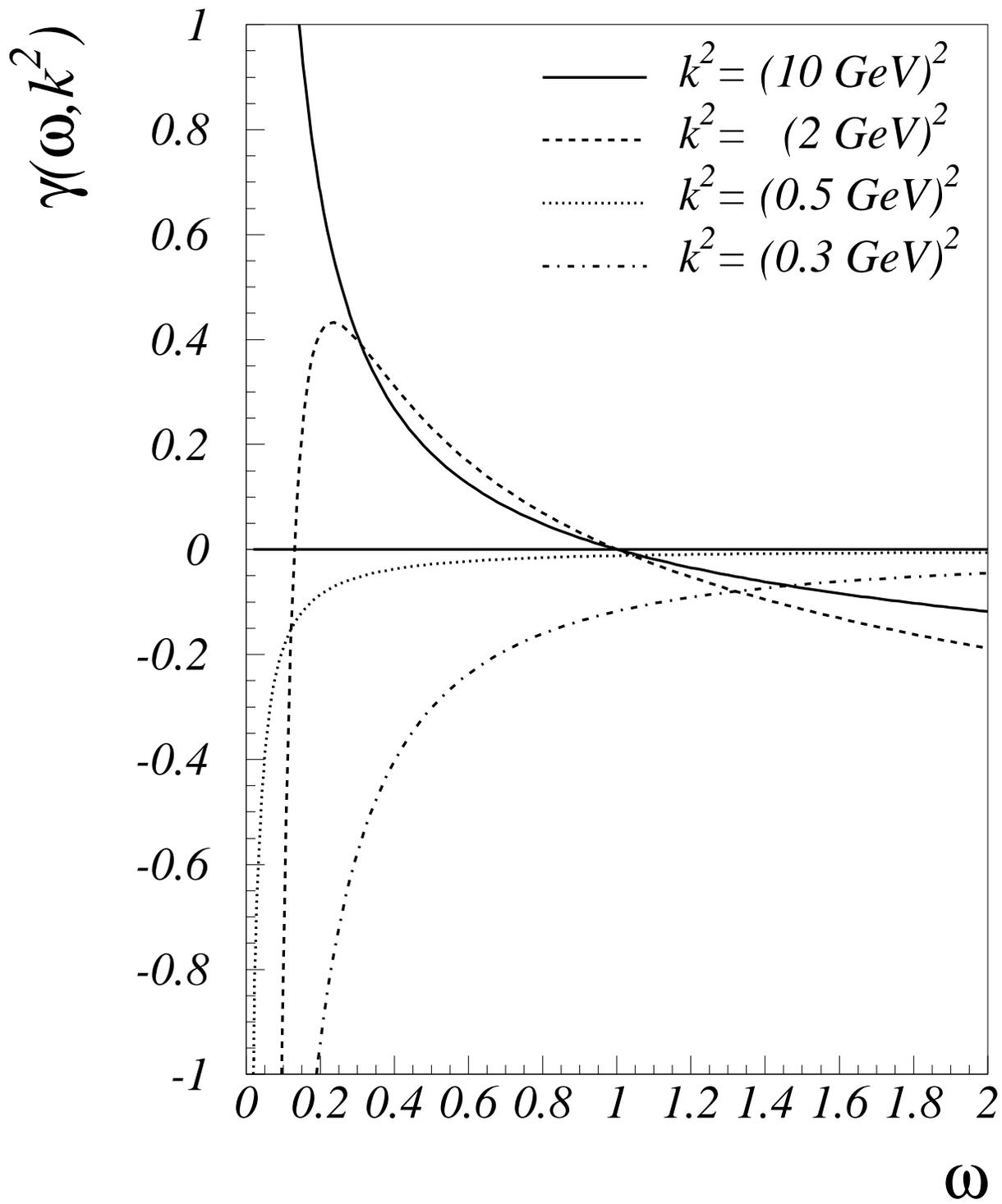
\noindent {\bf Figure 5:}
The anomalous dimension $\gamma (\omega,k_\perp^2)$ of eq. (\ref{andim})
versus $\omega$ for different values of $k_\perp^2$.
\bigskip

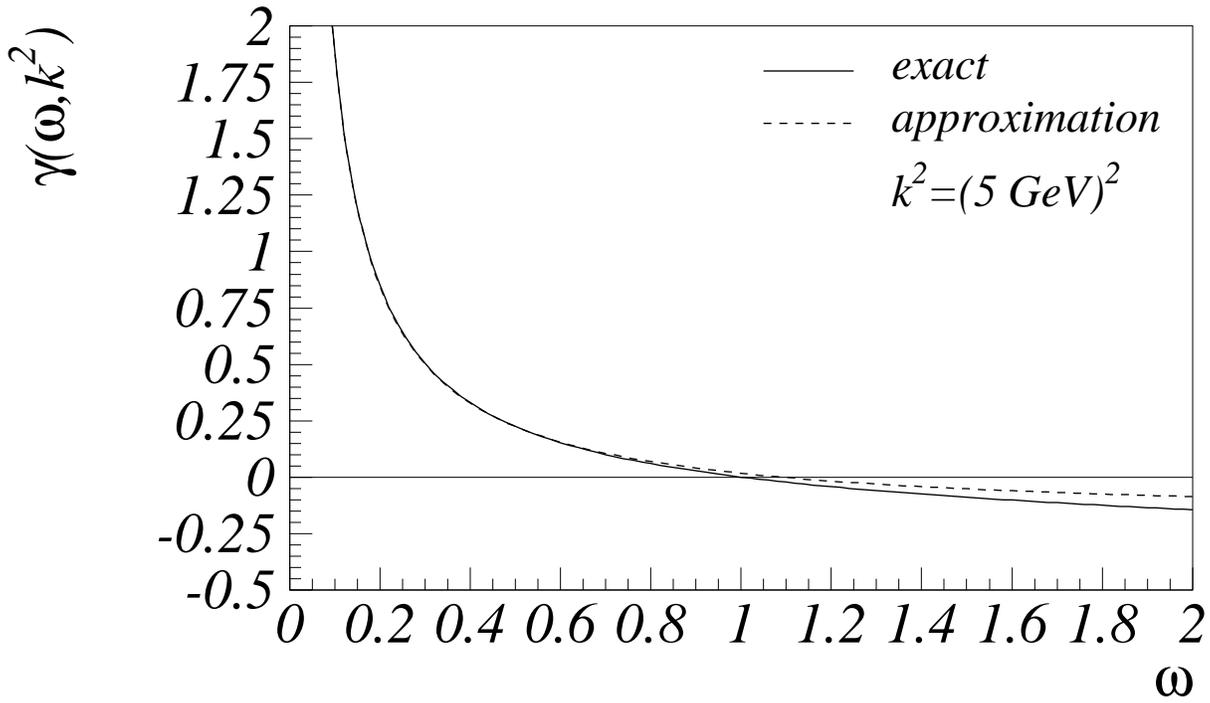
\noindent {\bf Figure 6:}
Comparison between exact expression for  the anomalous dimension
(\ref{andim}) and the approximations (\ref{ee7a}) [top] and
(\ref{ee7b}) [bottom].
\bigskip

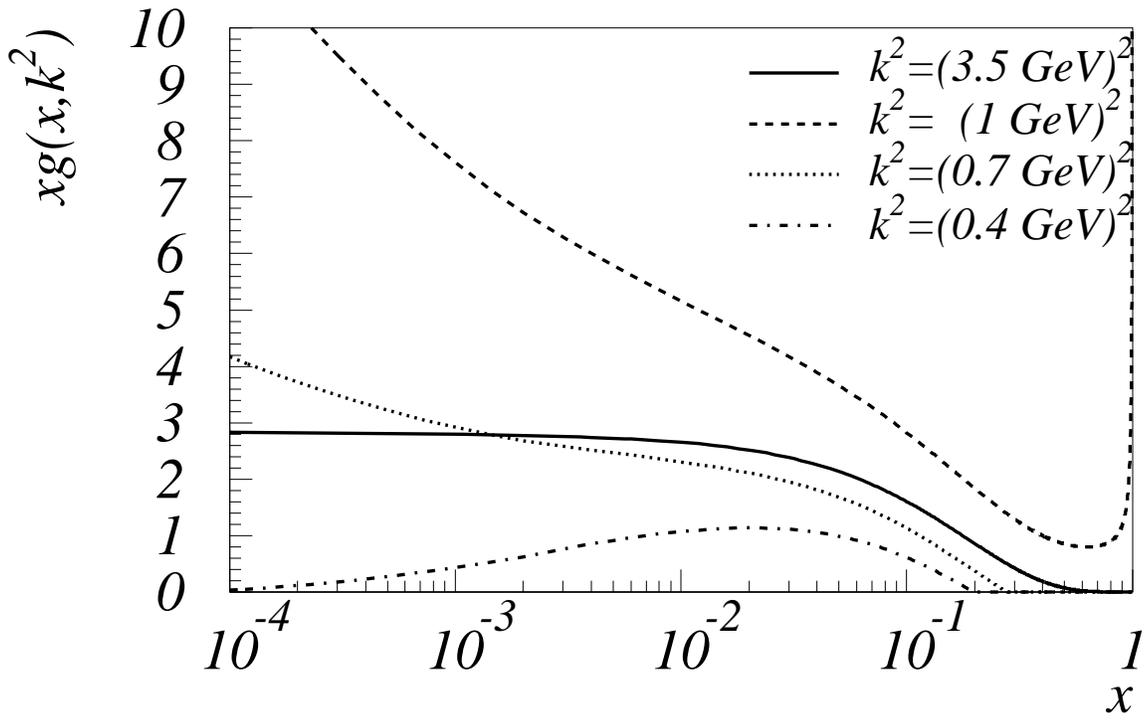
\noindent {\bf Figure 7:}
Behaviour of the distributions $xg(x,k_\perp^2)$ of (\ref{gsol1}) and
(\ref{gsol2})
for different values of $k_\perp^2$. Top part shows
result for a flat initial distribution
$xg(x,Q^2)=a (1-x)^c$ with $a=2.8$, $c=5.3$, and
bottom part for a Gaussian initial distribution
$xg(x,Q^2)=1/\sqrt{2\pi c^2} \exp[-(x-d)/(2c^2)]$ with $c=0.3$, $d=0.5$.
($Q=3.5$ GeV, $Q_0=1$ GeV, $\Lambda=0.23$ GeV).
\bigskip

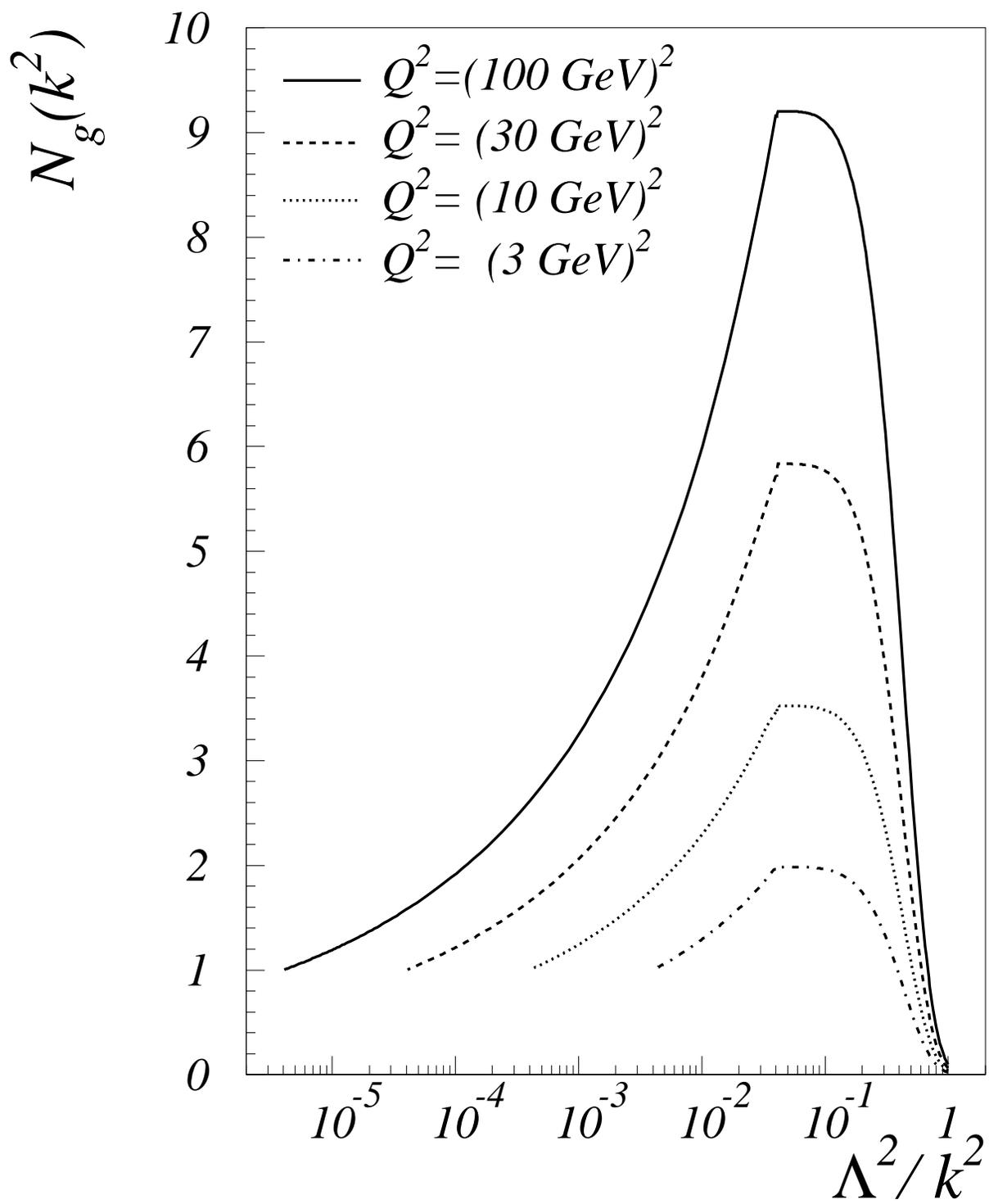
\noindent {\bf Figure 8:}
Evolution of the gluon multiplicity $N_g(k_\perp^2)$ from $Q^2$ down
to $\Lambda^2$. At first the gluons multiply, but at $k^2 \approx Q_0^2$
a condensation sets which is complete at $\Lambda^2$
($Q_0=1$ GeV, $\Lambda = 0.23$ GeV).
\bigskip

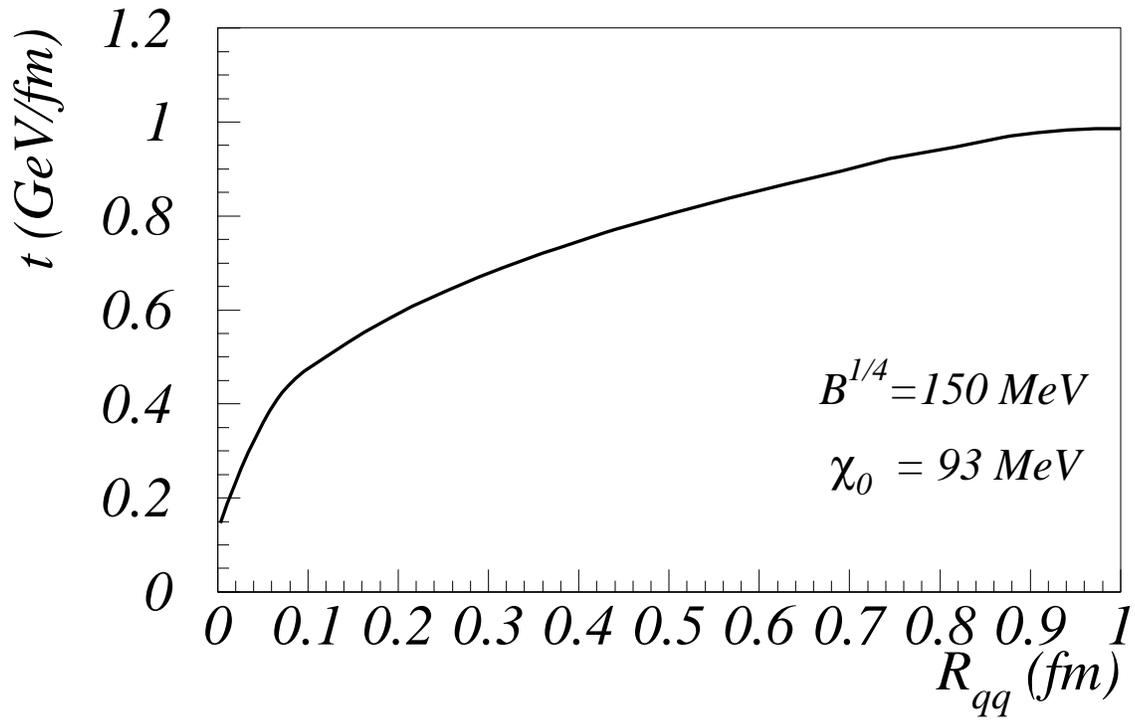
\noindent {\bf Figure 9:}
a) String tension $t$ versus separation $R_{q\bar q}$ of the $q\bar q$ pair,
and b) the solutions for  $\chi$ and $\kappa$ at $R_{q\bar q} =1$ fm versus
$r$ which is the radial coordinate perpendicular to $R_{q\bar q}$
($\chi_0=f_\pi$ and $B^{1/4}=150\,\hbox{MeV}$).
\bigskip

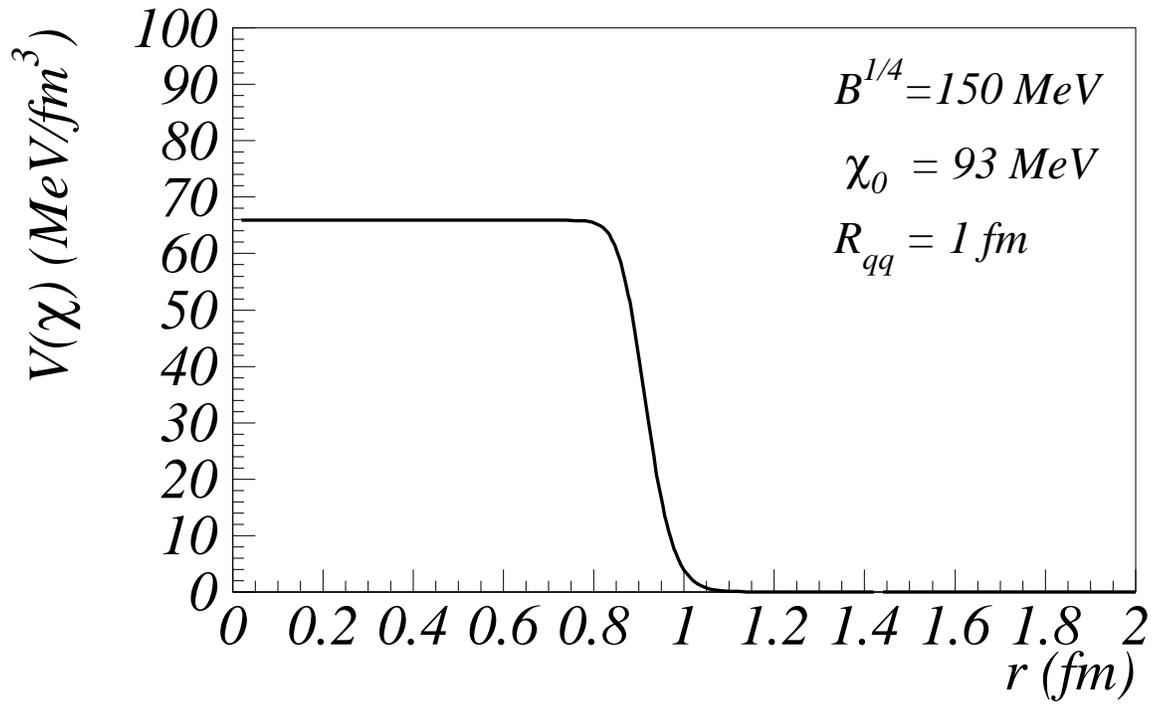
\noindent {\bf Figure 10:}
a) Form of the potential $V[\chi(r)]$
and b) the effective squared mass $M^2[\chi(r)]$ of the $\chi$ field,
both at $R_{q\bar q}=1$ fm.
($\chi_0=f_\pi$ and $B^{1/4}=150\,\hbox{MeV}$).

\vfill

\begin{thebibliography}{99}
\bibitem{colsys}
For the numerical computation, eq. (\ref{t2}) was decomposed into an
equivalent system of differential equations which was solved with
the program COLSYS, an algorithm for non-linear differential equations.
\end{thebibliography}
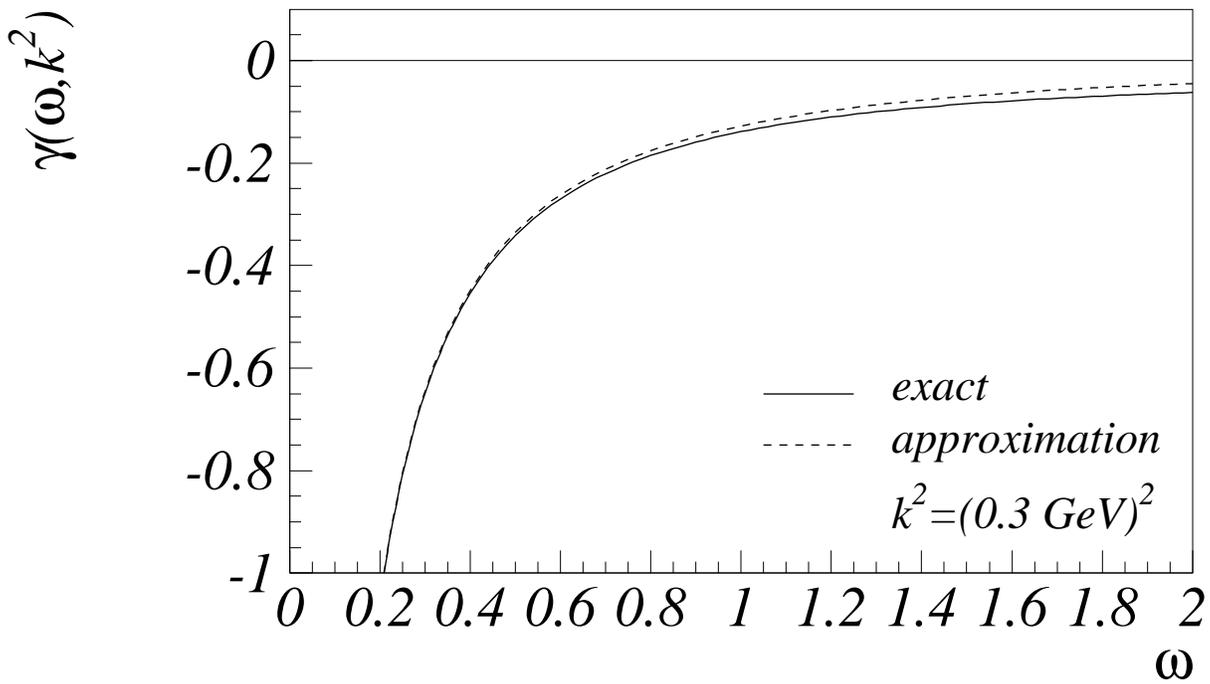
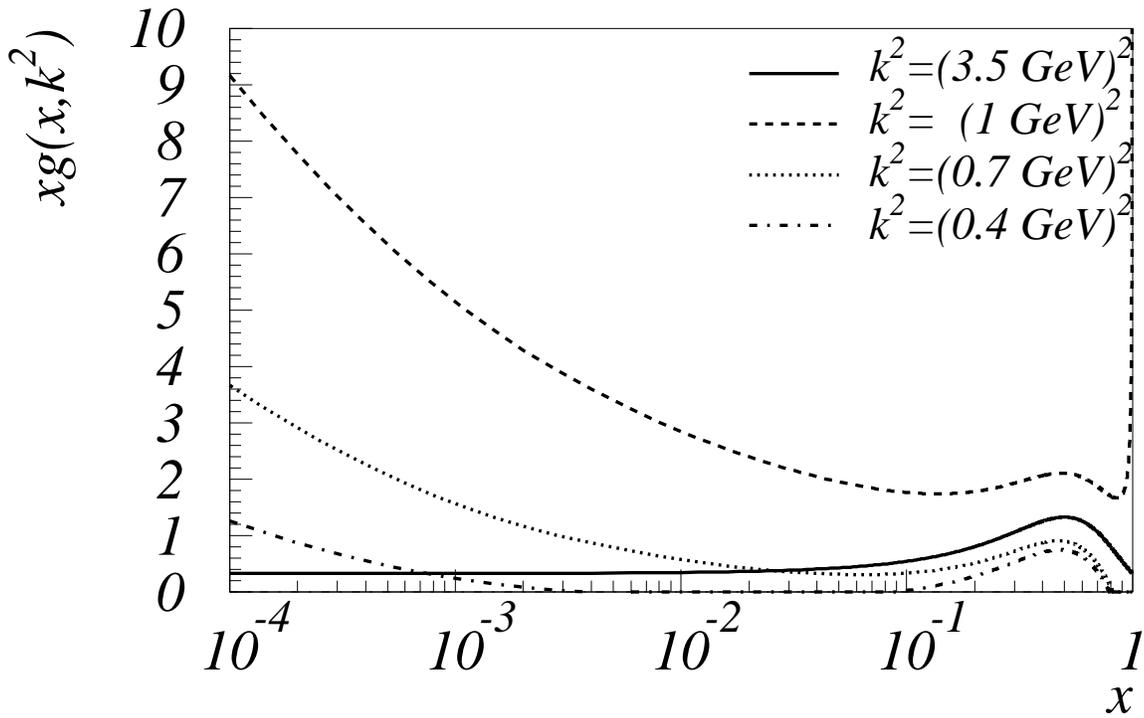
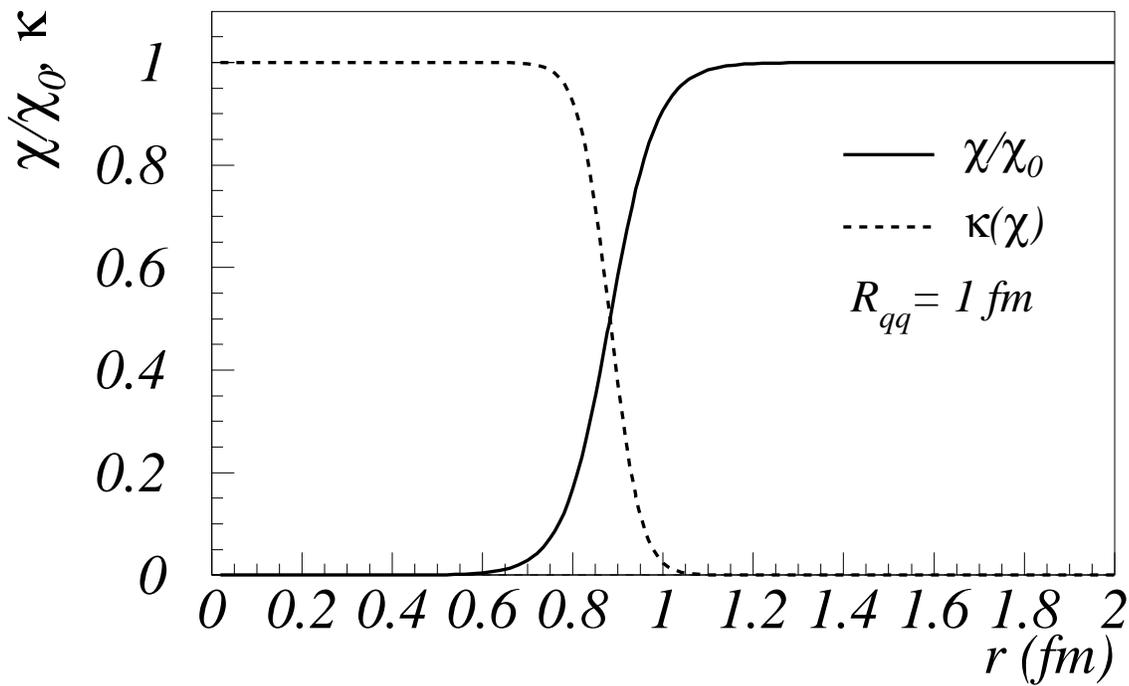
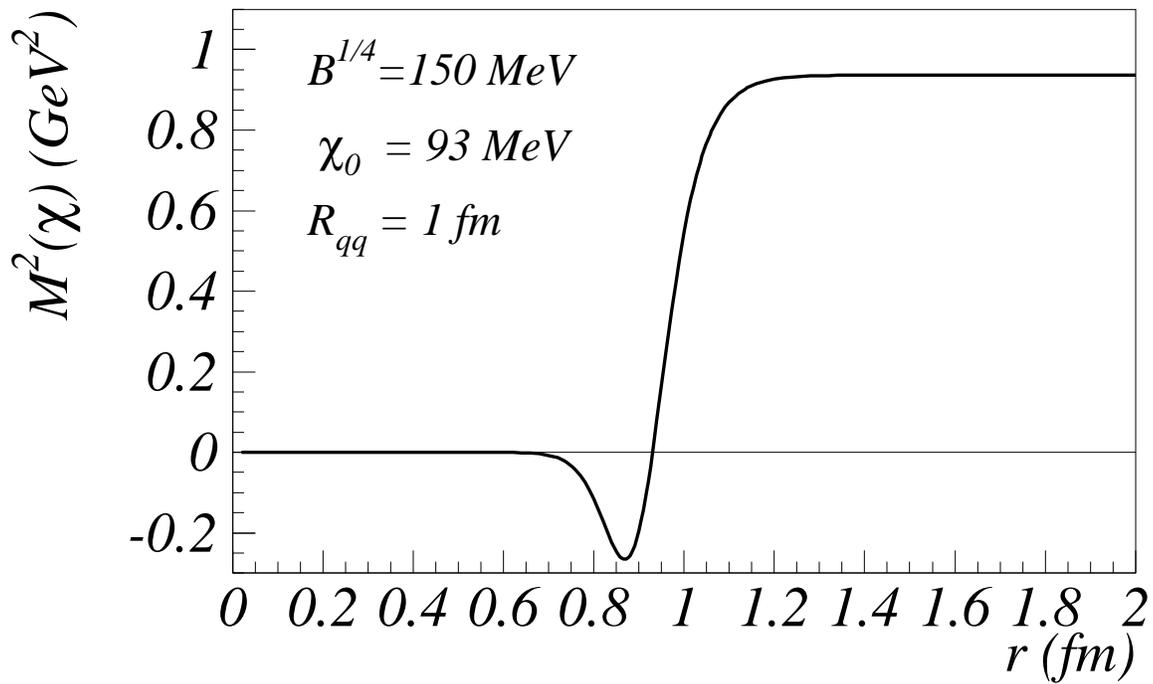
\end{document}